\documentclass[12pt]{article}

\usepackage{epsfig,epsf}

\begin{document}

\def\beq{\begin{equation}}

\def\eeq{\end{equation}}

\def\bea{\begin{eqnarray}}

\def\eea{\end{eqnarray}}

\def\eq#1{{Eq.~(\ref{#1})}}

\def\fig#1{{Fig.~\ref{#1}}}
             
\newcommand{\bas}{\bar{\alpha}_S}

\newcommand{\as}{\alpha_S} 

\newcommand{\bra}[1]{\langle #1 |}

\newcommand{\ket}[1]{|#1\rangle}

\newcommand{\bracket}[2]{\langle #1|#2\rangle}

\newcommand{\intp}[1]{\int \frac{d^4 #1}{(2\pi)^4}}

\newcommand{\mn}{{\mu\nu}}

\newcommand{\tr}{{\rm tr}}

\newcommand{\Tr}{{\rm Tr}}

\newcommand{\T} {\mbox{T}}

\newcommand{\braket}[2]{\langle #1|#2\rangle}

\newcommand{\ab}{\bar{\alpha}_S}

\newcommand{\x}{\vec{x}}
\newcommand{\y}{\vec{y}}
\newcommand{\z}{\vec{z}}
\newcommand{\rr}{\vec{r}}
\newcommand{\bb}{\vec{b}}
\newcommand{\bbb}{\vec{b}^{\,'}}
\newcommand{\xx}{\vec{x}^{\,'}}
\newcommand{\yy}{\vec{y}^{\,'}}
\newcommand{\zz}{\vec{z}^{\,'}}
\newcommand{\rrr}{\vec{r}^{\,'}}

\setcounter{secnumdepth}{7}

\setcounter{tocdepth}{7}

\parskip=\itemsep               

\setlength{\itemsep}{0pt}       

\setlength{\partopsep}{0pt}     

\setlength{\topsep}{0pt}        


\setlength{\textheight}{22cm}

\setlength{\textwidth}{174mm}

\setlength{\topmargin}{-1.5cm}





\newcommand{\lash}[1]{\not\! #1 \,}

\newcommand{\pd}{\partial}
\newcommand{\h}{\frac{1}{2}}

\newcommand{\g}{{\rm g}}

\newcommand{\el}{{\cal L}}

\newcommand{\A}{{\cal A}}

\newcommand{\Ka}{{\cal K}}

\newcommand{\al}{\alpha}

\newcommand{\be}{\beta}

\newcommand{\ep}{\varepsilon}

\newcommand{\ga}{\gamma}

\newcommand{\de}{\delta}

\newcommand{\dt}{\delta^{(2)}}

\newcommand{\De}{\Delta}

\newcommand{\et}{\eta}

\newcommand{\ka}{\vec{\kappa}}

\newcommand{\la}{\lambda}

\newcommand{\ph}{\varphi}

\newcommand{\si}{\sigma}

\newcommand{\ro}{\varrho}

\newcommand{\Ga}{\Gamma} 

\newcommand{\om}{\omega}

\newcommand{\La}{\Lambda}  

\newcommand{\tG}{\tilde{G}}

\newcommand{\lb}{\left(}
\newcommand{\rb}{\right)}

\renewcommand{\theequation}{\thesection.\arabic{equation}}



%

\def\ap#1#2#3{     {\it Ann. Phys. (NY) }{\bf #1} (19#2) #3}

\def\arnps#1#2#3{  {\it Ann. Rev. Nucl. Part. Sci. }{\bf #1} (19#2) #3}

\def\npb#1#2#3{    {\it Nucl. Phys. }{\bf B#1} (19#2) #3}

\def\plb#1#2#3{    {\it Phys. Lett. }{\bf B#1} (19#2) #3}

\def\prd#1#2#3{    {\it Phys. Rev. }{\bf D#1} (19#2) #3}

\def\prep#1#2#3{   {\it Phys. Rep. }{\bf #1} (19#2) #3}

\def\prl#1#2#3{    {\it Phys. Rev. Lett. }{\bf #1} (19#2) #3}

\def\ptp#1#2#3{    {\it Prog. Theor. Phys. }{\bf #1} (19#2) #3}

\def\rmp#1#2#3{    {\it Rev. Mod. Phys. }{\bf #1} (19#2) #3}

\def\zpc#1#2#3{    {\it Z. Phys. }{\bf C#1} (19#2) #3}

\def\mpla#1#2#3{   {\it Mod. Phys. Lett. }{\bf A#1} (19#2) #3}

\def\nc#1#2#3{     {\it Nuovo Cim. }{\bf #1} (19#2) #3}

\def\yf#1#2#3{     {\it Yad. Fiz. }{\bf #1} (19#2) #3}

\def\sjnp#1#2#3{   {\it Sov. J. Nucl. Phys. }{\bf #1} (19#2) #3}

\def\jetp#1#2#3{   {\it Sov. Phys. }{JETP }{\bf #1} (19#2) #3}

\def\jetpl#1#2#3{  {\it JETP Lett. }{\bf #1} (19#2) #3}


\def\ppsjnp#1#2#3{ {\it (Sov. J. Nucl. Phys. }{\bf #1} (19#2) #3}

\def\ppjetp#1#2#3{ {\it (Sov. Phys. JETP }{\bf #1} (19#2) #3}

\def\ppjetpl#1#2#3{{\it (JETP Lett. }{\bf #1} (19#2) #3} 

\def\zetf#1#2#3{   {\it Zh. ETF }{\bf #1}(19#2) #3}

\def\cmp#1#2#3{    {\it Comm. Math. Phys. }{\bf #1} (19#2) #3}

\def\cpc#1#2#3{    {\it Comp. Phys. Commun. }{\bf #1} (19#2) #3}

\def\dis#1#2{      {\it Dissertation, }{\sf #1 } 19#2}

\def\dip#1#2#3{    {\it Diplomarbeit, }{\sf #1 #2} 19#3 }

\def\ib#1#2#3{     {\it ibid. }{\bf #1} (19#2) #3}

\def\jpg#1#2#3{        {\it J. Phys}. {\bf G#1}#2#3}  

%


%


\def\thefootnote{\fnsymbol{footnote}} 

%

%

%


\noindent

\begin{flushright}

\parbox[t]{10em}{ \tt{TAUP 2795-05}\\

 \today 
}
\end{flushright}

\vspace{1cm}

\begin{center}

{\LARGE  \bf Towards a symmetric approach to high energy}\\
{\LARGE  \bf  evolution: generating functional with Pomeron loops}\\

\vskip1cm

{\large \bf ~E. ~Levin ${}^{a \,\ddagger}$ 
\footnotetext{${}^{\ddagger}$ \,\,Email:
leving@post.tau.ac.il, levin@mail.desy.de.}  
and ~M. ~Lublinsky ${}^{b \,\star}$ 

\footnotetext{${}^{\,\star}$ \,\,Email:
lublinsky@phys.uconn.edu, lublinm@mail.desy.de }}

\vskip1cm

{\it ${}^{a)}$\,\,\, HEP Department}\\
{\it School of Physics and Astronomy}\\
{\it Raymond and Beverly Sackler Faculty of Exact Science}\\
{\it Tel Aviv University, Tel Aviv, 69978, Israel}\\
\vskip0.3cm
{\it ${}^{b)}$\,\,\, Physics Department} \\
{\it University of Connecticut} \\
{\it U-3046, 2152 Hillside Rd., Storrs} \\
{\it CT-06269, USA}\\
\vskip0.3cm

\end{center}  

\bigskip

\begin{abstract}        
We derive an evolution equation  for the generating functional  
which accounts for processes
for  both gluon  emission and recombination. In terms of color dipoles, the kernel of this 
equation  describes
 evolution as a classical branching process with conserved probabilities. The
introduction of dipole recombination allows one  to obtained closed loops during the 
evolution, which should be interpreted as Pomeron loops of the BFKL Pomerons. 
In comparison with the emission, the dipole recombination is formally $1/N_c^2$ suppressed.
This suppression, nevertheless, is  compensated  at very high energies when
the scattering amplitude tends to its unitarity bound.
\end{abstract}

\newpage



\def\thefootnote{\arabic{footnote}} 

\section{Introduction}

As has been shown by Mueller \cite{MUDM},  the  high energy scattering in QCD
can be treated in the most economical  way in terms of color dipole degrees of freedom.
 In this approach, one considers a fast moving particle,   
as a  system  of colorless dipoles. The wavefunction of this system of dipoles can be 
found from the QCD generating functional  \cite{MUDM}. As was shown in Ref.  \cite{LL,LLB}
this functional obeys a  linear functional evolution equation (see also Ref. \cite{JANIK}).
This  linear functional evolution equation was derived in large 
$N_c$ approximation. For a small projectile, for which we can neglect nonlinear effects
associated with high dipole densities in its wavefunction, the functional evolution 
was shown to reproduce the dipole version of the Balitsky hierarchy \cite{B}
for the scattering amplitude. The latter reduces to the 
Balitsky-Kovchegov (BK) equation \cite{B,K}  if in addition we  assume  that the 
 low energy dipole interaction with the target has no correlations.

Though the BK equation has been widely used for phenomenology \cite{BKN}, it is clear 
starting from  the first papers on non-linear collective effects at high energies (see Refs. 
\cite{GLR,MUQI,MV}) that  simple non-linear evolution equations of the BK type could 
be correct only in a very limited kinematic range. At present there are several
approaches  allowing one to go beyond the BK equation. Balitsky \cite{B} has developed a 
Wilson
line approach which allows the  incorporation of both target correlations and $N_c$ 
corrections.
This method is  equivalent to the effective Lagrangian approach describing
the Color Glass Condensate and its derivative JIMWLK equation \cite{JIMWLK}. 

The methods above describe high energy evolution in a highly asymmetric manner: 
either the  projectile or target is always considered as a small perturbative probe.
Thus, the  constructed evolution contains a one way parton shower, in apparent violation of the 
$t$ channel
unitarity \cite{IM}. Is is  thus challenging to attempt to  restore the symmetry of the QCD 
evolution. 
If we are successful, we would be able to confront problems of high energy scattering of 
hadrons
or heavy nuclei in a reliable manner.

Several steps have been made  recently in attempt to formulate  a symmetric evolution.
Braun \cite{Braun} used the QCD triple Pomeron vertex \cite{BART}
 both for Pomeron splittings and mergings.  
Iancu and Mueller suggested a high energy factorization \cite{IM,KOLE,MUSO}, while
a statistical approach to  high energy scattering was proposed in Ref. \cite{IMM}.
In its turn,  Balitsky in Ref. \cite{BA} considered a symmetric scattering of two shock waves.
Another technique is due to Lipatov 
(e.g. \cite{LI}) who built an effective theory for reggeized gluons. Unfortunately, 
a relation between Lipatov`s theory and the other methods is so far not clear.
We want also to restore a symmetry without loosing the probabilistic interpretation of our 
results which leads to a simplest physical picture of the process  and a direct 
application for experimental observations.

A symmetric way to describe  high energy QCD does exist: it is 
the Reggeon technique \cite{GLR,MUQI,BART,NP}  based on interacting BFKL Pomerons \cite{BFKL}.
Many  elements of this technique are 
 known (see Refs. \cite{BART,NP}) and the main problem is to 
sum all Reggeon graphs. Past experience in summing  Reggeon diagrams does not look 
encouraging. However, a remarkable breakthrough was achieved in the last days of the Pomeron 
approach to  strong interactions: it turns out that the Reggeon calculus can be re-written in 
a probabilistic language. It was possible to formulate  equations
 for   probabilities to find a definite number of Pomerons at 
fixed 
rapidity \cite{GRPO,LEPO,BOPO}. The goal of this paper is to re-write the reggeon calculus of the BFKL 
Pomerons,  by extending our linear
functional approach based on dipole generating functional \cite{LL,LLB}. 
Colorless dipoles play two different roles in our approach. First, they are  partons for the BFKL 
Pomeron. This  role of the  dipoles is not related to the large $N_c$ approximation. 
Instead of defining a  probability to find  Pomerons 
we  search for  a probability to find a definite 
number of  dipoles at fixed rapidity. In this approach each vertex for splitting of 
one Pomeron into two Pomerons can be viewed as a decay of one dipole into two. Vise versa, 
merging of two Pomerons  is an annihilation process of two dipoles into one.  

The second role of the color dipoles is that at 
high energies they are good degrees of freedom. This fact
allows us to calculate  splitting and merging vertices. However, we need to 
stress  that the dipole model has been proven in the leading large $N_c$  approximation only. 
 The main assumption of this paper is that the dipole degrees of freedom can be in fact used for 
calculation of Pomeron vertices even beyond large $N_c$. 

Using this assumption
we derive, in addition
to previously known dipole splitting vertex $\Gamma(1\rightarrow2)$,  two new vertices. The first
one stands for the dipole recombination vertex $\Gamma(2\rightarrow 1)$. This vertex is derived
by computing a lowest order loop diagram and it is essentially the same as the
 triple BFKL Pomeron vertex  \cite{BART,NP}. The only difference is in the  normalization 
which allows us to use this vertex within the framework of the generating functional 
approach.

The second new vertex is $\Gamma(2\rightarrow3)$, which accounts for the  possibility 
of a dipole ``swing''. What we mean  is that with some probability two quarks of
a pair  of dipoles can exchange their antiquarks to form another pair  of dipoles.
Naturally, this process has $1/N_c^2$ suppression. It is the vertex $\Gamma(2\rightarrow3)$
that  correctly accounts for the Pomeron pairwise interaction  in the BKP equation \cite{BKP} and 
is absent in the usual form of the dipole evolution.

We observe several advantages of our approach based on the generating functional. 
\begin{itemize}
\item \quad Using the generating functional,  we can separate the structure of the 
wavefunction of the produced dipole at high energy  from rather complicated interaction of dipoles 
with the 
target at low energy. The latter are  subject to non-perturbative QCD  calculation and 
at the moment can be modeled only;
\item \quad Our approach is based on  colorless dipoles as the correct degrees of freedom 
at high energy. This fact allows us to have a natural description of  hard processes 
in perturbative QCD,  reproducing   linear evolution equations, 
such as DGLAP \cite{DGLAP} or/and BFKL \cite{BFKL}  equations;
\item \quad In our derivation of the linear functional equation we used a method which is 
closely related to the  probabilistic interpretation of the Reggeon Calculus (see Refs. 
\cite{GRPO,LEPO,BOPO}). In doing so we establish a clear correspondence between the color 
dipole approach to high energy scattering,  and Reggeon-like diagram technique providing a 
natural matching with  high energy phenomenology of soft processes based on Pomeron.
\end{itemize}

In the next section we describe the general formalism of the generating functional and its 
evolution equation taking into account both the emission of dipoles and their recombination.
We also derive the equations for the scattering amplitudes which solve the problem of
summation of the BFKL Pomeron loops. In section 3 and in the appendix we discuss the dipole 
vertices for the process of transition of two dipoles into one. Pomeron interactions
via  two to three dipole decay is a subject of Section 4.  Section 5 is devoted to 
study of  dynamic  correlations between dipoles.  In conclusions we summarize and discuss our results. 

\section{Dipole branching and Generating functional}

\subsection{Classical branching process and equation for probabilities.} 
We first consider a generic fast moving 
projectile whose wavefunction can be expended in a 
dipole basis. Note that contrary to many previous studies, we do not
restrict ourselves to a single dipole as a projectile.
\beq\label{db}
\Psi^{proj}\,=\,\sum_n\,\alpha_n\,|n\rangle
\eeq
Let us define a  probability density $P_n\,=\,\alpha_n^2$ to 
find  $n$ dipoles with coordinates $r_1, b_1$, $r_2, b_2$, $\dots r_i, 
b_i$, $\dots\,r_n, b_n$ 
and  rapidity $Y$ in the projectile wavefunction. 
$r_i$ and $b_i$ are correspondingly the  dipole's size
and impact parameter, both are two dimensional vectors.  
We define $P_n$ as a dimensionfull quantity which gives the  probability to find a dipole 
with the size $r_i$ (from $r_i$ to $r_i + dr_i$). The integral
\beq \label{P1}
\int \,\prod d^2 r_i \,\, P_n\left(Y\,-\,Y_0;\,r_1, b_1,\,r_2, b_2,
\dots r_n, b_n \right) \,\,=\,\,\Pi_n
\eeq
and  it gives the probability to find $n$-dipole with any sizes. This probability is conserved:  
$\sum_n \Pi_n =1$.

Suppose that  the following processes can occur as a result of one step in the evolution:
\begin{enumerate}
\item \quad The decay of a dipole with the size $r$ and impact parameter $b$ into two 
dipoles of the sizes $r'_1$ and $r'_2$ and impact parameters $b'_1$ and $b'_2$\footnote{We 
use notation $r_i,b_i$ for the initial state dipoles while $r'_l,b'_l$ for the final state 
ones.} , respectively:
\beq \label{G12}
 \Gamma( 1 \rightarrow 2)\equiv
\Gamma_{1\rightarrow2} \left( r,b \,\rightarrow\,r'_1,b'_1 \,+\,r'_2,b'_2 \right) 
\equiv \Gamma_{1\rightarrow2} \left(1;\,1'\,2' \right)  .
\eeq
\item \quad The annihilation of two dipoles with sizes $r_1$ and $r_2$ and impact 
parameters $b_1$ and $b_2$ into 
one dipole with the size $r'$ and impact parameter $b'$:
\beq \label{G21}
 \Gamma( 2 \rightarrow 1)\equiv
 \Gamma_{2\rightarrow1} \left( r_1,b_1\,+\, r_2,b_2\,\rightarrow\, r',b' \right) 
\equiv \Gamma_{2\rightarrow1}(1\,2;\,1') .
\eeq
\item \quad The process of interaction of two dipoles with sizes $r_1$ and $r_2$ with a 
creation of one additional dipole (not factorisable to $\Gamma( 1 \rightarrow 2)$ plus spectator): 
\beq \label{G23}
\Gamma(2 \rightarrow 3)\equiv  \Gamma_{2\rightarrow3} 
\left( r_1,b_1 \,+\,r_2,b_2 \,\rightarrow\,r'_1,b'_1\,+\, r'_2,b'_2\,+r'_3,b'_3\,\right) 
\equiv  \Gamma_{2\rightarrow3}(1\,2\,;\,1'\,2'\,3').
\eeq
\item \quad The annihilation process of three dipoles into two dipoles:
\beq \label{G32}
\Gamma(3 \rightarrow 2)\equiv  \Gamma_{3\rightarrow2} \left( r_1,b_1\,+\, r_2,b_2\,+r_3,b_3 
\,\rightarrow\,  r'_1,b'_1 \,+\,r'_2,b'_2 \right) \equiv  \Gamma_{3\rightarrow2}(1\,2\,3\,;\,1'\,2').
\eeq
\end{enumerate}
In the remaining of this Section and in the next one we will focus on the first two vertices only. We will
discuss $2\leftrightarrow 3$ transition in Section 4.

\begin{figure}[htbp]
\begin{center}
\epsfig{file= 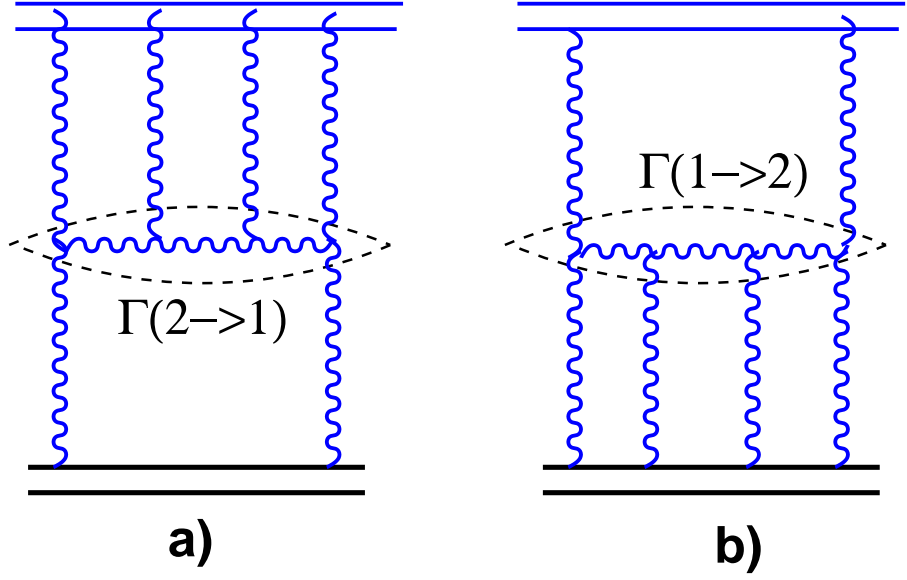,width=120mm}
\end{center}
\caption{\it The lowest order diagrams for $ \Gamma(2\,\rightarrow\, 1)$ (\fig{v12ba}-a)  
and for 
$\Gamma(1\,\rightarrow\, 2)$ (\fig{v12ba}-b)
.}
\label{v12ba}
\end{figure}

The equation for $P_n$ obeys the classical branching process:
 \beq \label{LEQPN}
\frac{\partial\,P_n\left(Y\,-\,Y_0;\,r_1, b_1,\,r_2, b_2, \dots r_n, b_n \right)}{
\partial\, Y}\,= 
\,-\, \sum^n_{i=1}\,\int \,d V_f  \, \Gamma_{1\rightarrow2}
 \left(i;\,1'\,2'\right)  P_n\left( \dots\,r_i,b_i 
 \,\dots \right)
\eeq
 $$+\,\sum^{n-1}_{i=1} \,\int dV_i \, \Gamma_{1\rightarrow2} \left(i';\,
i\,n \right) \, P_{n - 1}\left(
\dots r'_i,b'_i, \dots 
 \right)- \sum_{i \neq k} \int dV_f\, \Gamma_{2\rightarrow1}
 \left( i\,k \,;1'  \right)
\,\,P_n\left(\dots\,r_i, b_i, \dots 
\right)$$
$$+ \,\sum_{i \neq k} \int d V_i\, \Gamma_{2\rightarrow1}\left( i' \,k' ;
\,i \right)\,\,P_{n+1}\left(\dots\, r'_i,b'_i 
\,\dots\, r'_k,b'_k\dots
\right)$$

\eq{LEQPN} gives the general evolution for the probabilities $P_n$. \eq{LEQPN} must be 
supplemented by 
explicit expressions for the vertices $\Gamma(1 \,\rightarrow\, 2)$,
$\Gamma(2 \,\rightarrow\, 1)$.
By now, only the vertex $\Gamma(1 \,\rightarrow\, 2)$ has been 
calculated in Ref. \cite{MUDM}. On one hand, 
all other vertices are formally suppressed by $1/N_c$ and were 
omitted being considered as small. Moreover these vertices do not appear at all
within  the original formulation of the dipole model. On the other hand,   
 it should be stressed that the 
Feynman diagrams which correspond to these vertices have the same order of magnitude as far as 
the $N_c$ counting is concerned. For example,  two diagrams in \fig{v12ba} show the 
Born approximation for $\Gamma(1 \,\rightarrow\, 2)$ (\fig{v12ba}-b ) and for $ \Gamma(2 
\,\rightarrow\, 1)$  (\fig{v12ba}-a ). They have the same suppression in $N_c$ ($\approx 
1/N^4_c$) but in the diagram of \fig{v12ba}-b this suppression could be absorbed in the 
amplitude of the  interaction of two dipoles with the target while in \fig{v12ba}-a a factor 
$1/N^2_c$ has to be assigned to the vertex  $ \Gamma(2\,\rightarrow\, 1)$. 

\eq{LEQPN} has a very simple structure. 
For  every process of 
dipole splitting or merging   there are 
two terms: the first one, with the negative sign,  accounts for the probability $P_n$
to decreases due to splitting or merging  of one of  $n$ dipoles into  dipoles of arbitrary sizes 
and impact parameters;  the second term, with the positive sign, is responsible for an increase in 
probability to find $n$ dipoles due to the very same processes. The first 
term includes the vertex  $\Gamma(n \rightarrow k)$ integrated over the phase space of the 
final dipole: $\int \,dV_f\,\Gamma(n \rightarrow k) = \int \,\prod^k_l \,d^2 
r'_l\,d^2 b'_l\,\Gamma(n 
\rightarrow k)$. For the second term we need to integrate over the phase space of  initial 
dipoles: $\int \,dV_i\,\Gamma(n \rightarrow k) = \int \,\prod^n_l \,d^2
r_l\,d^2 b_l\,\Gamma(n
\rightarrow k)$.
Explicit expressions for the vertices $\Gamma$ will be given in the next section.

\subsection{Generating functional and linear functional evolution}
The hierarchy (\ref{LEQPN}) can be resolved by
introducing a generating functional $Z$
\beq \label{LD1}
Z\left(Y\,-\,Y_0;\,[u] \right)\,
\equiv\,\sum_{n=1}\,\int\,
P_n\left(Y\,-\,Y_0;\,r_1, b_1,\,r_2, b_2, \dots ,r_i, b_i, \dots ,r_n, b_n
 \right) \,
\prod^{n}_{i=1}\,u(r_i, b_i) \,d^2\,r_i\,d^2\,b_i
\eeq
where $u(r_i, b_i) \equiv u_i $ is an arbitrary function of $r_i$ and $b_i$. 
It  follows immediately from (\ref{LEQPN})
that the functional (\ref{LD1}) obeys the condition:
at $u\,=\,1$ 
\beq \label{LDIN2} 
Z\left(Y\,-\,Y_0;\,[u=1]\right)\,\,=\,\,1\,.
\eeq
The physical meaning of (\ref{LDIN2}) is that the sum over
all probabilities is one.

Multiplying \eq{LEQPN} by the product $\prod^n_{i=1}\,u_i$ 
and integrating over all $r_i$ and $b_i$,  we obtain the 
following linear equation for the generating functional:
\beq\label{Z}
\frac{\partial \,Z\,[\,u\,]}{
\partial \,Y}\,\,= \,\,\chi\,[\,u\,]\,\,Z\,[\,u\,]\,.
\eeq
Let us introduce the dipole collective coordinate $q$ with the 
integration measure $d^4\,q\,=\,d^2\,r\,d^2\,b$.
The evolution kernel $\chi$ is defined trough the operator vertices $V$
\begin{eqnarray}  \label{chi}
\chi[u]
&=& -\,\int\,d^4\,q\,V_{1\,\rightarrow \,1}\lb q;\,[u]\rb\
+\int\,d^4\,q_1\,d^4\,q_2\, V_{1\,\rightarrow \,2}
\lb q_1; q_2;\,
[u]\rb  \nonumber \\
&-&\int\,d^4q_1\,d^4q_2  V_{2\,\rightarrow \,2}\lb 
q_1;q_2;[u]\rb\,+\,\int\,d^4q' V_{2\,\rightarrow \,1}\lb q';[u] 
\rb \nonumber
\end{eqnarray}
The functional form of the vertices are related to $\Gamma$`s. 
\begin{eqnarray} \label{V11F} 
&&V_{1 \,\rightarrow\,1}\lb q,[u] \rb \,\,=\,\,\int\,\,d^4\,q_1\,d^4\,q_2\,
\Gamma_{1\rightarrow2} \left( q \,\rightarrow\,q_1 \,+\,q_2 \right) 
\,u(q)\,\frac{\de}{\de\,u(q)}.
 \\
&&V_{1 \rightarrow 2}(q'_1;\,q'_2;\,[u])\,\,
=\,\int d^4\,q\,\Gamma_{1\rightarrow2} \left( q \,\rightarrow\,q'_1 \,+\,q'_2 \right) 
\,u(q'_1)\,\,u(q'_2)\,
\frac{\delta}{\delta u(q)}\,.\label{V12F}
 \\
&&V_{2\,\rightarrow\,2}\lb q_1;\,q_2;[u] \rb\,\,=\,\,
\h\,\int\,\,d^4\,q\,\,\Gamma_{2\rightarrow1}\lb 
q_1\,+\,q_2\,\rightarrow\,q
\rb\,\,u(q_1)\,u(q_2)\,\frac{\de}{\de\,u(q_1)}\,\frac{\de}{\de\,u(q_2)}\,.\label{V22}\\
&&
V_{2\,\rightarrow\,1}\lb q;[u] \rb\,\,=\,\,
\h\,\int\,\,d^4\,q_1\,d^4\,q_2
\,\,\Gamma_{2\rightarrow1}\lb
q_1\,+\,q_2\,\rightarrow\,q
\rb\,\,u(q)\,\frac{\de}{\de\,u(q_1)}\,\frac{\de}{\de\,u(q_2)}\,.\label{V21}
\end{eqnarray}

The functional derivative with respect to $u(q)=u(r,b)$  plays a  role 
of an  annihilation operator for a dipole of the size $r$,  at the impact 
parameter $b$. 
The multiplication by $u(r,b)$ corresponds to
a creation operator for this dipole.

\eq{Z} exhibits a quantum mechanical-like structure with 
the operator $\chi$  viewed as a ``Hamiltonian'' of the evolution. In \eq{chi} $\chi$ is constructed
in terms of dipole creation and annihilation operators. 
 The evolution operator $\chi$ describes a
two dimensional  fully quantum  (nonlocal) field theory of interacting dipoles.

\subsection{Evolution of dipole densities}

The $n$-dipole densities in the projectile 
$\rho_n(r_1, b_1,\ldots\,,r_n, b_n)$
are defined as
\beq
\rho_n(r_1, b_1\,
\ldots\,,r_n, b_n)\,=\,\frac{1}{n!}\,\prod^n_{i =1}
\,\frac{\delta}{\delta
u_i } \,Z\left([u] \right)|_{u=1}
\eeq
Differentiating \eq{Z} $n$ times with respect to $u$ we can obtain a hierarchy of equations
for $\rho_n$
Refs. \cite{K,LL,LLB} and   rewrite \eq{LEQPN} in the form:
\begin{eqnarray} \label{rho}
&&\frac{\partial \rho_n \lb  r_1,b_1;\dots; r_n,b_n;Y \rb}{\partial\,Y} 
= \,\sum^{n - 1}_{i =1} \int d^4q\,
\Gamma_{1\,\rightarrow \,2} (q;\,q_i\,q_n)\,\rho_{n-1}\lb  \dots q \dots \rb \\
&&+
\, 2\sum_{i=1}^n\,\int\,d^4q'\,d^4q \,\Gamma_{1\rightarrow 2} (q';qq_i)\,
\rho_n\lb  \dots q'\dots \rb \,-
\sum_{i=1}^n \int d^4q_1'\,d^4q_2'
\,\Gamma_{1 \rightarrow 2} (q_i;q_1'q_2')\,\rho_n\lb  \dots q_i\dots \rb 
\nonumber\\
&&- \sum^n_{i,k,i\neq k}\,   \int \,d^4\,q\,
\Gamma_{2\,\rightarrow\,1} \lb q_i\,q_k;\,q \rb\, \rho_n\lb  \dots, 
q_i,\dots q_k\,\dots \rb   
\nonumber\\
&&-  \,2\,\sum_{i=1}^n \,\int d^4 q\, d^4 q' \Gamma_{2
\rightarrow 1} \lb q q_i;q'\rb\, \rho_{n+1}\lb \dots q_i \dots 
q\rb\, +\, \sum_{i=1}^n  \int \prod^2_{i=1} d^4q'_i\,
\Gamma_{2\rightarrow 1}\lb q'_1q'_2 ;q_i \rb \, \rho_{n+1}\lb q'_1 \dots 
q'_2\rb \nonumber
\end{eqnarray}
Eq. (\ref{rho}) presents a general structure with so far arbitrary vertices $\Gamma(1\rightarrow 2)$ 
and $\Gamma(2\rightarrow 1)$.
Explicit expressions  for $\Gamma$ will be presented in the next section.  
 The diagonal part of the evolution  due to the vertex 
$\Gamma(1\rightarrow 2)$ is the large $N_c$ limit of the BKP 
 equation \cite{BKP} in coordinate space. $n=1$ corresponds to 
 the evolution of the BFKL Pomeron.

The vertex $\Gamma(2\rightarrow 1)$  (\eq{ga21})  obeys the following
property $\int d^4 q\,\Gamma_{2\,\rightarrow\,1} \lb q_i\,q_k;\,q \rb\,=\,0$ \footnote{We are thankful to 
our referee who actually noticed   this property.}. Consequently, it is only the last term proportional to 
$\Gamma(2\rightarrow 1)$  who survives in  Eq. (\ref{rho}). 

\subsection{Scattering amplitude}
As was shown in  Refs. \cite{K,LLB}, the scattering amplitude is defined as a functional
\begin{eqnarray}  \label{N1}
&&N \lb Y;[\gamma_i] \rb =- \sum^{\infty}_{n =1}
\,\frac{(-1)^n}{n!}\,\int
\,\gamma_n(r_1,b_1; \dots r_n,b_n;Y_0) \,\,\prod^n_{i=1}\,\frac{\de}{\de 
u_i} Z\lb Y,[u_i] \rb|_{u_i =1}  d^2 r_i\,d^2b_i\,\nonumber\\
 &&=- \sum^{\infty}_{n =1}(-1)^n \,\int\,\gamma_n(r_1,b_1; 
\dots r_n,b_n;Y_0)\,\rho_n \lb r_1,b_1;\dots r_n,b_n;Y_0 \rb \,\prod_{i=1}^n\, d^2 r_i\,d^2b_i\, .
\end{eqnarray}
The amplitude for simultaneous 
scattering of $n$ dipoles off the target  is denoted by $\gamma_n$.
It has to be specified at the lowest rapidity ($Y_0$).
Using the ansatz $ N\lb (Y,[\gamma_i] \rb = N \lb \gamma_1(Y),\gamma_2(Y) \,\dots 
\gamma_i(Y) 
\dots \rb$ and having
\beq \label{N4}
\frac{\partial\, N}{\partial\,Y}
=\, \sum^{\infty}_{n =1}\,\int\,\prod^n_{i =1}\,d^2\,r_i\,d^2
b_i\,\,\frac{\de\,N}{\de \gamma_n}\,\,\frac{\partial\,\gamma_n}{\partial\,Y}
\eeq
we can recast the hierarchy of equations (\ref{rho})  into the Balitsky-type chain 
for the scattering amplitudes (see Ref. \cite{LLB})
\begin{eqnarray} \label{N5}
&&\frac{\partial\, \gamma_n\lb r_1,b_1,\dots, r_n,b_n\rb}{\partial\,Y}\,\,\,\,= \\
&=&
 2\,\sum_{i=1}^n\,\int\,d^4q'\,d^4q \,\Gamma_{1\,\rightarrow \,2} (q_i;\,q\,q')\,
\gamma_n\lb  \dots q'\dots \rb \,
 -\,   \sum_{i=1}^n \,\int d^4q_1'\,d^4q_2'\,
\Gamma_{1 \,\rightarrow\,2} (q_i;\,q_1'\,q_2')\,\gamma_n\lb  \dots, q_i\dots \rb 
\nonumber \\ 
 &-&\sum^{n - 1}_{i =1} \int d^4q\,d^4q'
\Gamma_{1\,\rightarrow \,2} (q_i;\,q\,q')\,\gamma_{n+1}\lb  \dots q \dots q'\rb 
+\,\sum_{i\ne j}^n\,\int  d^4\,q\,
\Gamma_{2\,\rightarrow\,1}\lb q_i\,q_j ;\,q \rb  
\gamma_{n-1}\lb 
 q_i \dots  q_j \dots q\rb \nonumber \\
&-& 2\,\sum_{i=1}^n \int d^4 q\, d^4 q' \Gamma_{2
\rightarrow 1} \lb q q_i;q'\rb \, \gamma_{n-1}\lb \dots q_i \dots 
q\rb
\,+ \,\sum^n_{i,k,i\neq k}\,   \int d^4q\,
\Gamma_{2\rightarrow 1} \lb q_i q_k; q \rb\,\gamma_n\lb  \dots 
q_i\dots q_k \dots \rb
\nonumber 
\end{eqnarray}

 The great advantage of \eq{N1} 
is the fact that this equation allows 
us to take into consideration in the most economic way the  interaction 
of low energy dipoles with 
the target. For example, assuming  $\rho_1 (Y_0) = \dt \lb \vec{r} \,-\,\vec{r}_1 \rb \dt 
\lb 
\vec{b} 
\,-\,\vec{b}_1 \rb$ while $ \rho_{n>1}=0$, we obtain that the total amplitude of a single dipole 
scattering equals to
\beq \label{N7}
N\lb r,b;Y \rb \,\,\,=\,\,\,\gamma_1\lb r,b;Y \rb\,.
\eeq
If we assume the projectile be built out of  two dipoles with \\
 $ \rho_2(Y_0)\,\,=\,\, \dt \lb \vec{r} \,-\,\vec{r}_1 \rb \dt \lb 
\vec{b}\,-\,\vec{b}_1 
\rb
\dt \lb \vec{r}^{\,'} \,-\,\vec{r}_2 \rb \dt \lb \vec{b}^{\,'}\,-\,\vec{b}_2 \rb$ and
$\rho_1=0 $, $\rho_{n>2}=0$ then
\beq \label{N8}
N\lb r,b;r',b'; Y \rb \,\,\,=\,\,\,\gamma_2\lb r,b;r'b';Y \rb\,.
\eeq
\eq{N5} is an evolution hierarchy for dipole amplitudes. Apparently it involves loop processes.
The equation is most general for postulated vertices. 
It has a very similar structure 
as suggested by  Iancu and Triantafyllopoulos in Ref. \cite{IT} (Eqs.(6.6) 
and (6.7) of this paper). In the following section we will present
explicit expressions for all $\Gamma$.
 The exact vertex
$\Gamma( 2 \,\rightarrow \,1)$   found by us (see \eq{ga21})   does not seem to  coincide with
the vertex suggested in Ref. \cite{IT} in any kinematic region\footnote{Our vertex does coincide
with the one found by the authors of \cite{IT} in their  paper \cite{IT1}, which appeared after our
preprint started to circulate.}. 

In fact, the last two terms in \eq{N5}  vanish for the vertex  $\Gamma( 2 \,\rightarrow \,1)$
given by \eq{ga21}. Nevertheless we prefer to keep these terms explicitly in the hierarchy \eq{N5}. 
The only reason behind keeping them is that in practical applications one may attempt to approximate or
simplify the vertex $\Gamma( 2 \,\rightarrow \,1)$. For an approximate vertex the last two terms might not vanish.
Diagrammatically these terms are part of the $4 \rightarrow 4$ reggeized gluon transition
which  may contribute  for some kinematics
where the underlying probability conservation is important.

\section{Dipole vertices}
\begin{boldmath}
\subsection{$\Gamma(1\,\, \rightarrow\,\,2)$}
\end{boldmath}

The vertex for the decay of one dipole into two has been derived in Ref. \cite{MUDM}:
\begin{eqnarray}\label{V12}
&&\Gamma_{1\rightarrow2}\lb r,b\,\, \rightarrow\,\,r'_1,b'_2\,+\,r'_2,b'_2 \rb\,\,\,= \\
&&\bas\frac{r^2}{r'^2_1\,\,r'^2_2}\,\dt\lb \vec{r} \,-\,\vec{r}^{\,'}_1 \,-\,\vec{r}^{\,'}_2 \rb\,\,
\dt\lb \vec{b}'_1 \,-\,\vec{b} \,+\,\h\vec{r}'_2 \rb  \dt\lb \vec{b}'_2 \,-\,\vec{b} 
\,-\,\h\vec{r}'_1 \rb \,.\nonumber
\end{eqnarray}
As has been discussed, this vertex leads to reduction of the probability to 
find $n$ dipoles due to 
decay into two dipoles of arbitrary sizes. This reduction is related to
\beq \label{V12R}
\bas \,\omega(r)\,\,\equiv\,\,\int\,dV'\,\Gamma_{1\rightarrow2}\lb r,b\,\, \rightarrow\,\,r'_1,b'_2\,+\,r'_2,b'_2 
\rb\,\,=\,\,\frac{\bas}{2\,\pi}\,\int_{\rho}\,\frac{r^2}{r'^2\,(\vec{r}\,-\,\vec{r}')^2}\,d^2\,r'
\eeq
where $\bas = \as N_c/\pi$ and $\rho$ is the infrared cutoff.
The growth term is proportional to
\beq \label{V12G}
\int\,dV\,\Gamma_{1\rightarrow2}\lb r,b\,\, \rightarrow\,\,r'_1,b'_2\,+\,r'_2,b'_2\rb\,
=\,\int\,d^2r\,d^2b\,\,\Gamma_{1\rightarrow2}\lb r,b\,\, \rightarrow\,\,r'_1,b'_2\,+\,r'_2,b'_2\rb
\eeq
$$
\,\,=\,\,\frac{\bas}{2\,\pi}\,\frac{(\vec{r}'_1 \,+\,\vec{r}'_2)^2}{r'^2_1\,r'^2_2}\,\dt \lb 
\vec{b}'_1\,-\,\vec{b}'_2 \,-\,\h (\vec{r}'_1 \,+\,\vec{r}'_2) \rb\,.
$$
So far, most of the discussions of dipole evolutions were bounded to the vertex 
$\Gamma(1\rightarrow 2)$  as all other vertices 
are $1/N^2_c$ suppressed  and were considered as  small corrections. 
Below we will include several new vertices, which   are of the order  $1/N^2_c$.
These additional vertices give important contributions in the deep saturation 
region (see Refs. \cite{GLR,IM,KOLE,MUSO,IMM} for more detailed discussion of this subject) and
hence need to be accounted for. Though we do not pretend to be able to accommodate all of the $N_c$
corrections using dipole degrees of freedom, we believe the contributions we aim to
include are dominant at high energies.

\begin{boldmath}
\subsection{$\Gamma(2\,\, \rightarrow\,\,1)$}
\end{boldmath}

\begin{figure}[ht]
    \begin{center}
        \includegraphics[width=0.90\textwidth]{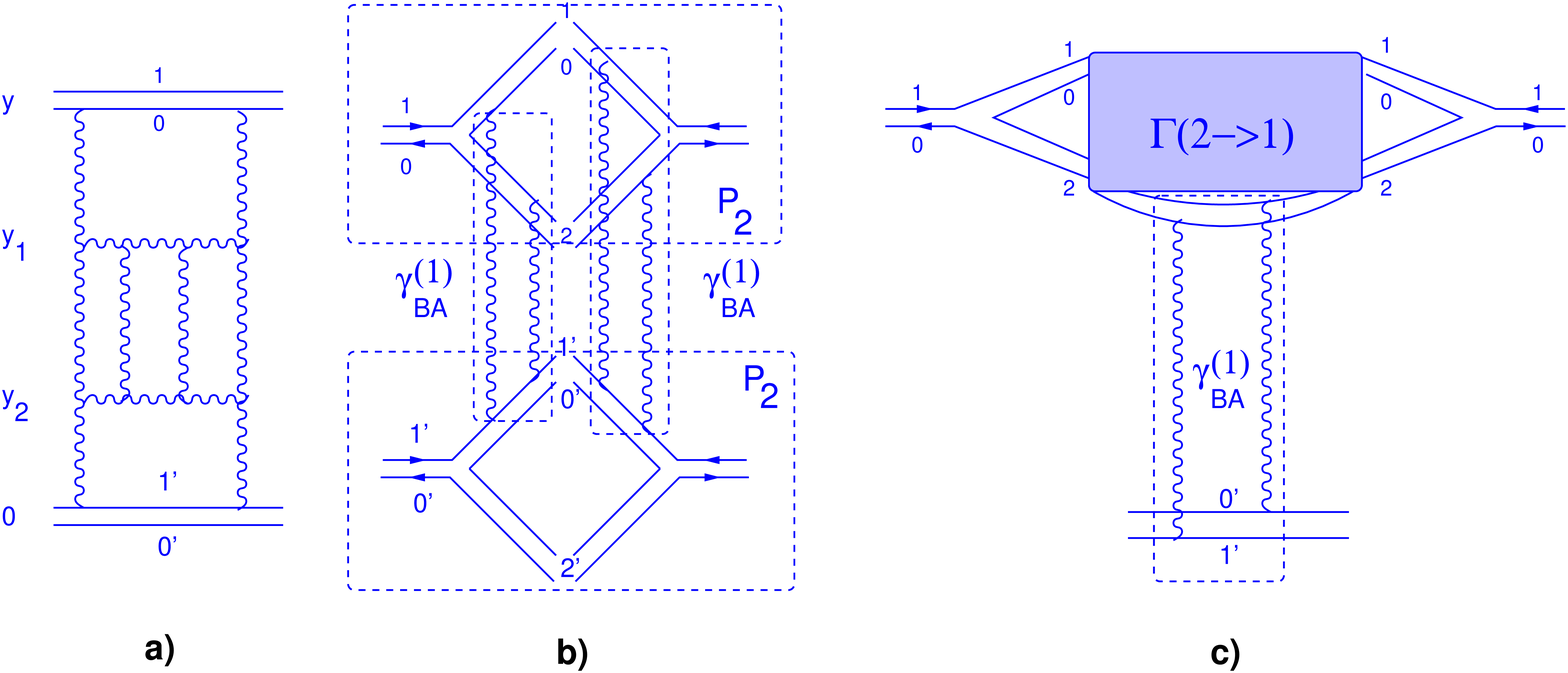}
        \caption{\it The first enhanced diagram and  the vertex of annihilation of 
two dipoles 
into one dipole. }
\label{v21di}
    \end{center}
\end{figure}

To find the vertex  $\Gamma( 2 \to 1)$   we analyze the 
 first enhanced diagram shown in 
\fig{v21di}-a. As was shown in Refs. \cite{IM,KOLE} the expression for this diagram is  
\begin{eqnarray}
\label{G211}
&&\int\,d^2\,r_1\,d^2\,b_1\,\,d^2\,r_2\,d^2\,b_2\, 
\,d^2\,r'_1\,d^2\,b'_1\,d^2\,r'_2\,d^2\,b'_2\,P_2\lb r'_1\,b'_1;r'_2,b'_2\rb \nonumber\\
&&\,\,\,\,\,\,\,\,\,\,\,\,\,\,\,\,\,\,\,\,\,\,\,P_2\lb 
r_1,b_1;r_2,b_2\rb
\,\gamma^{(1)}_{BA}\lb r_1,r'_1,\vec{b}_1\,-\,\vec{b}^{\,'}_1\rb\,\,\gamma^{(1)}_{BA}\lb 
r_2,r'_2,\vec{b}_2\,-\,\vec{b}^{\,'}_2\rb 
\end{eqnarray}
where $\gamma^{(1)}_{BA} $ is a dipole-dipole elastic
 scattering  amplitude due to exchange of 
two gluons. The expression for this amplitude is well known
 (see Refs. \cite{BAA,LI})
\beq \label{G21G}
\gamma^{(1)}_{BA}\lb r_1,r'_1;\vec{b}_1\,-\,\vec{b}^{\,'}_1\rb\,\,=\,\,
\frac{\bar{\alpha}^2_s}{32\,N_c^2}\,\,
\,\ln^2 \lb \frac{(\vec{\Delta b}\,+\,\vec{R})^2\,(\vec{\Delta b}\,-\,\vec{R})^2}{(\vec{\Delta 
b}\,+\,\vec{\Sigma})^2\,(\vec{\Delta b}\,-\,\vec{\Sigma})^2} \rb\,\,\equiv\,\,
\eeq
$$
\equiv\,\,\frac{\bar{\alpha}^2_S}{32\,N^2_c}\,\ln^2 \lb \frac{(\vec{x}_1 - 
\vec{x}^{\,'}_1)^2\,(\vec{y}_1 - 
\vec{y}^{\,'}_1)^2}{(\vec{x}_1 - \vec{y}^{\,'}_1)^2\,\,(\vec{y}_1 - \vec{x}^{\,'}_1)^2}\rb
$$
where $\vec{\Sigma}\,=\,\h( \vec{r}_1 \,+\,\vec{r}^{\,'}_1)$,  $\vec{R}\,=\,\h( \vec{r}_1 
\,-\,\vec{r}^{\,'}_1)$,  $ \vec{\Delta b}\,=\,\vec{b}_1\,-\,\vec{b}^{\,'}_1$; and $\vec{r}_1\,=\,
\vec{x}_1 - \vec{y}_1$, $ \vec{r}^{\,'}_1\,=\,\vec{x}^{\,'}_1 - \vec{y}^{\,'}_1$.

\eq{G211} follows from the fact that we can view this diagram in the following way. The upper 
dipole ($x_{10}$ in \fig{v21di}) evolves with normal vertex $\Gamma (1 \rightarrow 2) $ in the 
rapidity interval $y - y_1$ (see \fig{v21di}-a) while the low dipole ($x_{1'0'}$ in \fig{v21di}) 
also evolves but in rapidity interval $y_2$. As the result of these two evolutions there are 
 two dipoles  with the sizes $r_1 = x_{20}$ and $r_2=x_{12}$ at the rapidity $y_1$ and two dipoles
with sizes $r'_1 = x_{2'0'}$ and $r_2'=x_{1'2'}$ at the rapidity $y_2$. Each pair elastically 
rescatters  by the
exchange of two gluons leading to \eq{G211}.

Alternatively, \eq{LEQPN} gives another expression for the same diagram
\beq \label{G212}
\int\,d^2\,r_1\,d^2\,b_1\,d^2\,r_2\,d^2\,b_2\, 
\,d^2r\,d^2\,b\,P_2 \lb r'_1\,b'_1 \, r'_2\,b'_2\rb\,
\Gamma_{2\rightarrow1}\lb
r_1,b_1\,+\,r_2,b_2 \rightarrow r,b \rb
\gamma^{(1)}_{BA}\lb r,r';\vec{b}\,-\,\vec{b}^{\,'}\rb
\eeq
where $r' = x_{1'0'}$ in \fig{v21di}. Comparing \eq{G211} and \eq{G212} we obtain
the following equation for $\Gamma_{ 2\rightarrow1}\lb
r_1,b_1\,+\,r_2,b_2 \rightarrow r,b \rb$:
\beq \label{G213}
\int\,d^2\,r\,d^2\,b\, \Gamma_{2\rightarrow1}\lb
r_1,b_1\,+\,r_2,b_2 \rightarrow r,b \rb
\,\gamma^{(1)}_{BA}\lb r,r',\vec{b}\,-\,\vec{b}^{\,'}\rb\,\,=\,\,
\eeq
$$
=\,\,\,\int d^2\,r'_1\,d^2\,b'_1\,d^2\,r'_2\,d^2\,b'_2\,\Gamma_{1\rightarrow2}\lb r',b' \rightarrow 
r'_1,b'_1 + r'_2,b'_2\rb\,
\,\gamma^{(1)}_{BA}\lb r_1,r'_1;\vec{b}_1\,-\,\vec{b}^{\,'}_1\rb\,\,\gamma^{(1)}_{BA}\lb
r_2,r'_2,\vec{b}_2\,-\,\vec{b}^{\,'}_2\rb\,.
$$
\eq{G213} is the basic equation from which the vertex $\Gamma(2\rightarrow 1)$ can be extracted.
To this goal we need to invert \eq{G213} by acting on both sides of it by an operator inverse 
to $\gamma_{BA}$ in operator sense. Fortunately, this operator is known to be a product of two Laplacians:
\beq \label{DDBA}
\Delta_x\,\Delta_y\,\gamma^{(1)}_{BA} \lb x,y; x',y' \rb\,\,=\,\,\as^2 \left(
\delta^{(2)}\lb  {x} -  {x}^{\,'}\rb \delta^{(2)}\lb  {y} -  {y}^{\,'}\rb
\,+\,\delta^{(2)}\lb  {x} -  {y}^{\,'}\rb \delta^{(2)}\lb  {y} - 
 {x}^{\,'}\rb \right)
\eeq
with $x=b+r/2$ and $y=b-r/2$ being the coordinates of quark and antiquark in the dipole $(r,b)$.
For the vertex  $\Gamma(2\rightarrow 1)$ we finally obtain 
\begin{eqnarray}\label{ga21}
\Gamma_{2\rightarrow1}\lb
r_1,b_1\,+\,r_2,b_2 \rightarrow r,b \rb&=&\frac{1}{\alpha_S^2}\,\Delta_x\,\Delta_y\,
\int d^2\,r'_1\,d^2\,b'_1\,d^2\,r'_2,d^2\,b'_2\,\Gamma_{ 1\rightarrow2}\lb r,b \rightarrow 
r'_1,b'_1 + r'_2,b'_2\rb\,\times\nonumber\\
&\times &
\,\gamma^{(1)}_{BA}\lb r_1,r'_1,\vec{b}_1\,-\,\vec{b}^{\,'}_1\rb\,\,\gamma^{(1)}_{BA}\lb
r_2,r'_2;\vec{b}_2\,-\,\vec{b}^{\,'}_2\rb\,.
\end{eqnarray}
In Appendix A we present a method for  evaluation of  the expression (\ref{ga21}).
We arrive at the result given by \eq{A13}. 
Since the general expression is rather  complicated 
we  consider now simplified estimates valid in several different kinematic 
regions.

In the region where $\Delta b \,\ll\,r'$ and $r \,\ll\,r'$ \eq{G21G} leads to a simple expression
\beq \label{G21GS}
\gamma^{(1)}_{BA}\lb r,r';\Delta b \rb\,\,=\,\,\frac{2\bas^2}{N^2_c}\,\frac{r^2}{r^{\,'2}}
\eeq
Using \eq{G21GS} we can rewrite \eq{G213} in the simple form if we are looking for the 
contribution in the following kinematic region:
$$ r_1 
\,\,\approx\,\, 
r_2\,\gg\,r'_1\,\,\approx\,\,\,r'_2\,\,\,\,\gg\,\,\,r' 
\,\,\,\,\,\mbox{and}\,\,\,\,r\,\,\gg\,\,r'$$
In this kinematic region using \eq{V12} and \eq{LEQPN} we obtain that  the r.h.s. of \eq{G213} is 
equal to
\beq \label{G214}
\mbox{r.h.s.\,\,of\,\,\eq{G213}}\,\,\,=\,\,\,\,\bas\,\,\left(\frac{\bas}{N^2_c} 
\right)^2\,\int\,\,d^2 
r'_1 \,d^2\,r'_2\,d^2\,b'_1\,d^2 b'_2\,\,
\eeq
$$
\lb \gamma^{(1)}_{BA}\lb 
r_1,r'_1;\vec{b}_1 - \vec{b}^{\,'}_1 \rb\,=\,\,\frac{r^{\,'2}_1}{r^2_1} \rb
\,\,
\lb \gamma^{(1)}_{BA}\lb
r_2,r'_2;\vec{b}_2 - \vec{b}^{\,'}_2 \rb\,=\,\,\frac{r^{\,'2}_2}{r^2_2} \rb
$$
$$
\,\frac{r'^2}{r^{\,'2}_1\,\,r^{\,'2}_2}\,\,\dt \lb \vec{r}^{\,'}\,- 
\,\vec{r}^{\,'}_1\,-\,\vec{r}^{\,'}_2 \rb\,\dt  \lb \vec{b}^{\,'}_1\,-\,\vec{b} \,+\,\h 
\vec{r}^{\,'}_2 \rb \,\dt \lb \vec{b}^{\,'}_2\,-\,\vec{b}\,-\,\h \vec{r}^{\,'}_1 \rb \,\,=\,\,
$$
$$
=\,\,\,\bas\,\,\lb
\frac{\bas^2}{N^2_c}\,\rb^2\,\int \,\,d^2 
\,r'_1\,\,\frac{r'^2}{r^{\,'2}_1\,\,r^{\,'2}_2}\,\frac{r^{\,'2}_1\,
\,r^{\,'2}_2}{r^2_1\,r^2_2}\,\,
=\,\,\,\,\,\bas\,\,\lb
\frac{\,\bas^2}{N^2_c}\,\rb^2\,\frac{r'^2}{r^2_1}
$$
The l.h.s of \eq{G213}  has a form
\beq \label{G215}
\mbox{l.h.s.\,\,of\,\,\eq{G213}}\,\,=
\eeq
$$
\,\,\frac{32\,\bas^2}{N^2_c}\, \int\,\,d^2 
r\,d^2\,b\,\Gamma_{2\rightarrow1}\lb r_1,b_1\,+\, 
r_2,b_2\,\rightarrow\,r,b \rb \lb \gamma^{(1)}_{BA}\lb
r,r';\vec{b} - \vec{b}^{\,'} \rb\,=\,\,\frac{r^{2}}{r^{\,'2}} \rb\,=\,
$$
$$
=\,\,\frac{32\,\bas^2}{N^2_c}\pi \int^{r^2_1}_{r'^2}\,\,r'^2\,d r^2\, \Gamma_{ 2\rightarrow1}
\lb r_1,b_1\,+\,
r_2,b_2\,\rightarrow\,r,b \rb \,
$$
In the last equation we used the fact that in \eq{G21G}
 the typical $b$ is about the size of the 
large dipole ($b \,\approx\,r'$).
Substituting  \eq{G214} and \eq{G215} in \eq{G213} we obtain 
$\Gamma \lb 2 \, \rightarrow\,1 \rb$ 
in the form
\beq \label{G216}
\Gamma_{2\rightarrow1}\lb r_1,b_1\,+\,r_2,b_2\,\rightarrow\,r,b 
\rb\,\,=\,\,\frac{(4\,\bas)^3}{N^2_c}\,\, \frac{1}{r^4_1}\,\,
\eeq

Repeating the same calculation but in a different kinematic region:
$$
 r_1 
\,\approx\,r_2\,\ll\,r'_1\,\,\approx\,r'_2\,\,\,\mbox{and} 
\,\,\,r'_1\,\,\approx\,r'_2\,\,\gg\,\,r'\,\,\,\mbox{while} \,\,\,r\,\,\ll\,\,r'
$$
we obtain 
\beq \label{G217}
\Gamma_{ 2\rightarrow1}\lb r_1,b_1\,+\,r_2,b_2\,\rightarrow\,r,b
\rb\,\,=\,\,\frac{(4\,\bas)^3}{N^2_c}\,\, \frac{r^2_1\,r^2_2}{r^8}
\eeq
The full expression for $\Gamma \lb 2 \,\rightarrow\,1 \rb $ is rather complicated as
can be 
seen from \eq{A13}. In corresponding limits it reproduces 
\eq{G216} and \eq{G217}.

\begin{boldmath}
\section{ Pomeron interaction: $2\,\rightarrow 3$ transition vertex}
\end{boldmath}

In this section we further extend the dipole model by introducing an additional splitting vertex,
$\Gamma(2\rightarrow 3)$. Our main goal here is to account for Pomeron pairwise interactions
via exchange of a single gluon. For the first time this process was included in the double 
log approximation of pQCD in Refs. \cite{BALE,LALE}. 
It is also most naturally included in the BKP equation \cite{BKP} 
providing $1/N_c^2$ corrections to the  dipole evolution discussed above.

The inclusion of the above Pomeron interactions in terms of dipole degrees of freedom is
 not a  straightforward task. 
We face two problems here. First, the contribution we are looking for is a process in which a gluon 
is emitted (in the amplitude)  by one dipole and then reabsorbed (in the conjugate amplitude) 
by another dipole. This is an
interference contribution, which does not admit a probabilistic interpretation. 
The exact expression for $N_c$ corrections to the dipole evolution 
known\footnote{We thank Yu. Kovchegov who drew our attention to 
Ref. \cite{MC} after our preprint started to circulate.} 
from Refs. \cite{MC,KL2}
can, nevertheless, be projected onto dipole degrees of freedom. By introducing the vertex 
$\Gamma(2\rightarrow3)$ we take into account only the diagonal contributions
factorisable in terms of dipoles. We trust that the rest of the $N_c$ corrections  
contribute to multi gluon $t$-channel states only. The latter, $n$-gluon states
 are known to have intercepts smaller than that of $n/2$ Pomerons
 and thus could be ignored at high energies.

The second problem is in the fact that dipoles are natural degrees of freedom in the large $N_c$ limit
only. Beyond large $N_c$, the dipole basis (\ref{db}) is overcomplete. In particular, a single space configuration
of two pairs of quarks and  antiquarks can be counted twice as two different pairs of dipoles 
(provided all quarks are mutually in a color singlet state).  As a result of working with overcomplete
basis there will be a nontrivial overlap between probabilities to find a different number of dipoles.

\begin{figure}[ht]
    \begin{center}
        \includegraphics[width=0.90\textwidth]{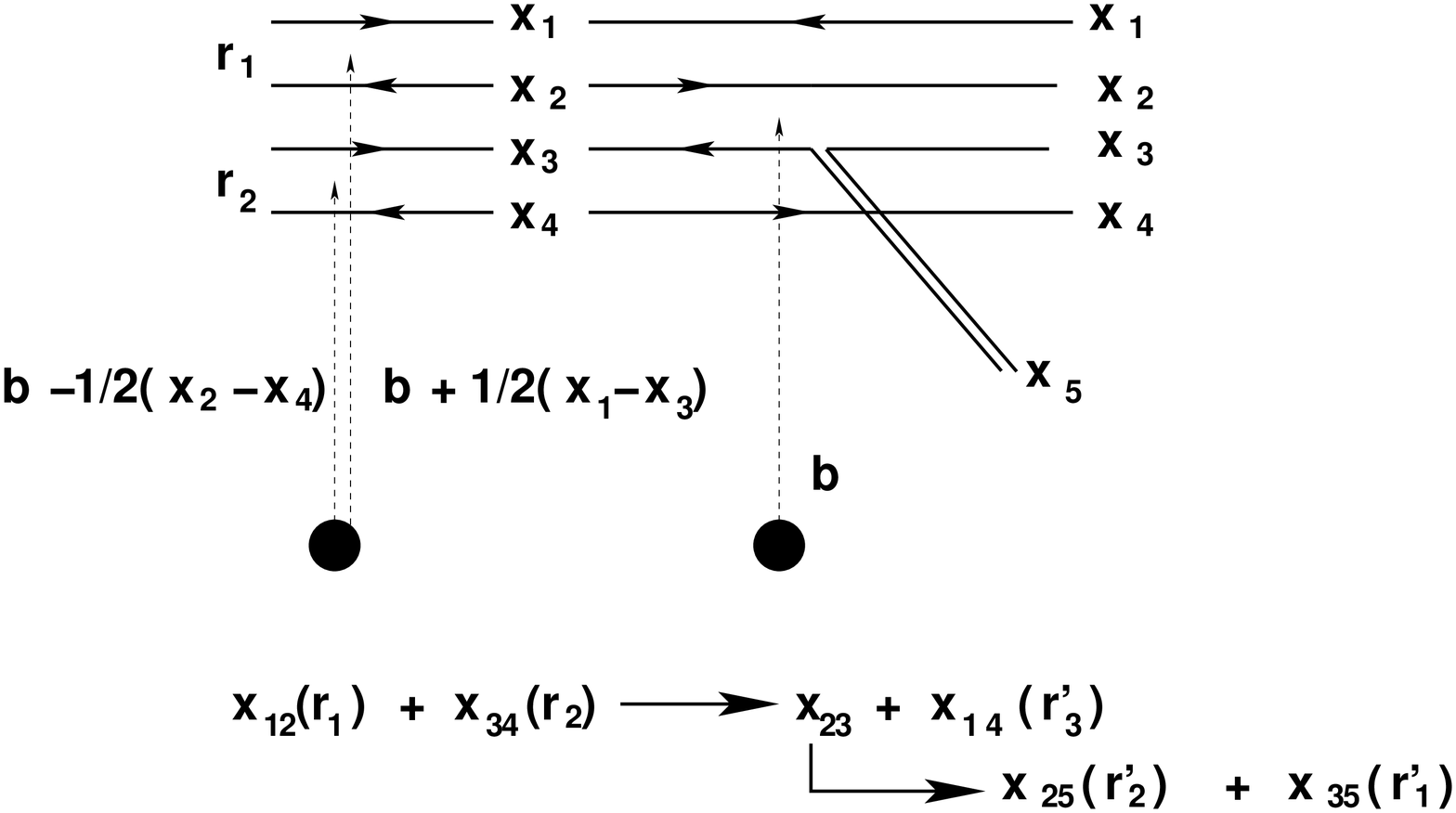}
        \caption{\it The process $ x_{12} + x_{34} \,\rightarrow\, x_{14} + x_{35} + x_{45}$. }
\label{23}
    \end{center}
\end{figure}

Having sorted the above problems out, we propose the following vertex $\chi^{2\rightarrow3}$
to be added to the dipole evolution kernel $\chi$: 
\begin{eqnarray}\label{chi23}
\chi^{2\rightarrow3}\,&=&\,\int_{x_1,\, x_2 ,\,x_3\,x_4,\,x_5}
\Gamma_{2\rightarrow 3} \left(x_2,x_3,x_5\right)\,\,\times \\
&\times&\left[1\,-\,u(x_1,x_4)\right]\,
\left[ u(x_2,x_3)\,- \,u(x_2,x_5)\,u(x_5,x_3) \right]\,\frac{\de}{\de\,u(q_1)}\,\frac{\de}{\de\,u(q_2)}\,.\nonumber
\end{eqnarray}
The operator (\ref{chi23}) describes the following process. First, it annihilates two dipoles
 $q_1=(x_1,\,x_2)$ and $q_2=(x_3,\,x_4)$ (\fig{23}).  
Then the dipoles are regrouped into $14$ (spectator) and $23$. The latter 
subsequently decays through the usual $1\rightarrow2$ dipole splitting process
($ u(x_1,x_4)\,( u(x_2,x_3)$ \\ $- u(x_1,x_4)\,u(x_2,x_5)\,u(x_5,x_3) )$ term in the operator).
The vertex $\Gamma(2\rightarrow 3)$ has the usual dipole splitting form ($\Gamma(1\rightarrow 2)$)
suppressed by a factor $N_c^2$:
$$\Gamma_{2\rightarrow 3}(x,\,y,\,z)\,=\,\frac{\bar\alpha_s}{2\,\pi\,N_c^2}\,
\frac{(x\,-\,y)^2}{(x\,-\,z)^2(y\,-\,z)^2}$$

In (\ref{chi23}) we have also subtracted a term with the spectator $u(x_1,x_4)$ set to unity.
This subtraction is needed to remove the double counting: the decay of a single dipole ($23$)
has been already accounted for in the normal $1\to 2$ dipole evolution. This subtraction can be 
also thought of as originating from the overcomplete basis we are dealing with. 
 We will find below
that this subtraction is crucial to prevent the operator (\ref{chi23}) from generating 
Pomeron loops at the level  of scattering amplitudes
\footnote{We have missed this subtraction in the first preprint version of this paper.
We are most thankful to our colleagues Ian Balitsky, Jochen Bartels, Al Mueller, Yura Kovchegov, 
Alex Kovner, and the referee whose criticism helped us to solve the problem.}.

For the evolution of the dipole densities $\rho_n$ the operator $\chi^{2\rightarrow 3}$
generates the following contribution (for $n\ge 2$) 
\begin{eqnarray}\label{rho23}
\frac{\partial \rho_n \lb  q_1;\dots; q_n;Y \rb}{\partial\,Y} &=&
 \sum_{i,j,k=1,\,i\ne j\ne k}^n \,\Gamma_{2\rightarrow3}(x_j,y_k,y_j)
\,\rho_{n-1}(\dots x_i,y_k\dots x_j,y_i\dots)\,\delta^2(x_k\,-\,y_j) \nonumber \\
&+&\, \sum_{i,j=1,\,i\ne j}^n\int d^2 y\,\Gamma_{2\rightarrow3}(x_j,y,y_j)\,
\rho_n(\dots x_i,y\dots x_j,y_i\dots) \nonumber \\
&-&\, \sum_{i,j=1,\,i\ne j}^n\int d^2 z\,\Gamma_{2\rightarrow3}(x_j,y_j,z)\,
\rho_n(\dots x_i,y_j\dots x_j,y_i\dots) 
\end{eqnarray}

The evolution of the scattering amplitudes receives additional terms (for $n\ge 2$):
\begin{eqnarray}\label{gamma23}
&&\frac{\partial \gamma_n \lb  q_1;\dots; q_n;Y \rb}{\partial\,Y}\,= 
 \sum_{i,j,\,i\ne j}^n \,\int d^2z\,\Gamma_{2\rightarrow3}(x_j,y_i,z) 
\left[\gamma_n(\dots x_i,y_j\dots x_j,z\dots)\,+\,\right. \\
&&\gamma_n(\dots x_i,y_j\dots z,y_i\dots) 
-\left.\gamma_n(\dots x_i,y_j\dots x_j,y_i\dots)\,-\,
\gamma_{n+1}(\dots x_i,y_j\dots x_j,z\dots z,y_i\dots)\right] \nonumber
\end{eqnarray}
Eq. (\ref{gamma23}) supplemented by the usual dipole evolution generated by the vertex $\Gamma(1\rightarrow 2)$
is believed to be a very good approximation to the Balitsky-JIMWLK evolution. The advantage of 
our formulation is that it is given entirely in terms of dipole degrees of freedom. We will demonstrate
in the following section that the above evolution happens to coincide with the one  found in Ref. 
\cite{BLV} by analyzing $N_c$ corrections arising from the QCD triple Pomeron vertex \cite{BART}.

Finally let us comment about  $3 \,\rightarrow 2$ transition vertex.
The process of $3 \,\rightarrow 2$ has $1/N^4_c$ compared to the leading $1\,\rightarrow 2$.
 Indeed, it progresses in two stages: the first one is the annihilation 
of two dipoles into one.  Such a process has $1/N^2_c$ suppression. Then two remaining dipoles 
``swing'' quarks (see \fig{23}) and this  has  an additional $1/N_c^2$ 
suppression. Therefore, $\Gamma(3\,\, \rightarrow\,\,2)$ is of the order of $1/N^4_c$ 
and will be neglected.

\begin{boldmath}
\section{$N_c$ correlations due to $2\,\rightarrow\,3$ vertex}
\end{boldmath}
Let us combine \eq{N5} and \eq{gamma23} but neglect  the vertex $\Gamma(2\,\rightarrow\,1)$. 
We would like to find a procedure
which would allow the equations entering the hierarchy to decouple from each other. In case of
original Balitsky`s hierarchy this was achieved by assuming absence of target correlations which means
substitution of the Kovchegov`s factorization \cite{K}:
 \beq \label{N9}
\gamma_n \lb r_1,b_1 \,\dots,\,r_n,b_n;Y \rb \,\,\,=\,\,\prod^n_{i=1}\,\,\gamma_i \lb r_i,b_i;Y
\rb\,.
\eeq
The whole hierarchy respected the factorization leaving only one single equation (BK) unresolved.

Since we have dynamical correlations, the hierarchy of  \eq{N5}+\eq{gamma23} obviously does not  admit the 
factorization of  \eq{N9}.
An intuitive solution would be to introduce pairwise correlations which would hopefully reduce the hierarchy
to two coupled equations.
The natural generalization of \eq{N9} is to  introduce two-dipole
correlation $C \lb r_i,b_i; r_k,b_k;Y\rb $ in the form
\beq \label{N10}
\gamma_n \lb r_1,b_1 \,\dots,\,r_n,b_n;Y \rb \,\,\,=
\eeq
$$
\,\,\prod^n_{i=1}\,\, \gamma_i \lb r_i,b_i;Y\rb\,\,\,+\,\,\sum^n_{i=1,k=1,i \neq 
k}\prod^n_{l=1,l\neq i,l \neq k}\,\,C \lb r_i,b_i; r_k,b_k;Y\rb \,\,\gamma_i \lb 
r_l,b_l;Y\rb
$$
\eq{N10} can be written compactly by introducing the operator $Q$, such that
 \beq \label{N100}
\gamma_n \lb r_1,b_1 \,\dots,\,r_n,b_n;Y \rb \,\,\,= Q[\gamma_1]\,\,\prod_i^n \,\gamma_1(r_i,\, b_i)\,,
\eeq
with
$$Q[\gamma_1]\,\,\equiv\,\,Exp\,\lb \int\,d^4q_1\,d^4q_2\, C \lb q_1; \,q_2;\,Y\rb\,
\frac{\delta}{\delta\,\gamma_1(q_1)}\,\frac{\delta}{\delta\,\gamma_1(q_2)}\rb\,.
$$
We have checked that, though the introduction of correlations in the form \eq{N10} is very plausible idea,
this ansatz does not make the system of hierarchy equations to decouple. Nevertheless, 
we can try to estimate the influence of the new 
vertex by taking into account the correlations between dipoles in  perturbative way 
considering them small. To this goal we will focus on the first two equations of \eq{N5}+\eq{gamma23}
which will allow us to determine the evolution law for $\gamma_1$ and $C$. 
Introducing $K$ as the usual dipole kernel
\beq \label{K}
K\lb  {x}, {y};  {z}\rb \,\,\,=\,\,\frac{(  {x} -  {y})^2}{( {x} - 
 {z})^2\,\,( {x} \,-\, {z})^2}\,
\eeq
the equations for $\gamma_{1,2}$ read
\begin{eqnarray}
&&\frac{\partial\, \gamma_1\lb x,y\rb}{\partial\,Y}\,=\,  
\bas\int_z\,K\lb 
 {x}, {y}; 
 {z} \rb\,\lb   {-\,\gamma_1}\lb  {x},  {y}\rb
+\,\gamma_1\lb  {x},  {z} \rb\,+\,\gamma_1 \lb  {y},  {z} 
\rb -\,\gamma_2 \lb  {x},  {z};  {y} ,  {z} \rb  \rb
\, \,\,\,\,\,\,\,\,\,\,\,\,\,\,\nonumber \\ \label{NG1} \\  
&&\frac{\partial\, \gamma_2\lb x_1,y_1; x_2,y_2\rb}{\partial\,Y}\,= \, \nonumber \\
&& \, \bas\,\int_z\,
\sum_{i,j=1,\,i\ne j}^2\,
K\lb 
 {x_i}, {y_i}; 
 {z} \rb\,\left[ 
\gamma_2\lb  {x_i},  {z};x_j,y_j \rb\,+\,\gamma_2 \lb  {y_i},  {z};x_j,y_j 
\rb {-\,\gamma_2}\lb  {x_i},{y_i}; x_j, y_j\rb
\right] \nonumber\\
&&+\,\,\frac{\bas}{N_c^2}\,\int_z 
 \sum_{i,j=1,\,i\ne j}^2 \,K(x_j,y_i;z) \,
\left[\gamma_2(x_i,y_j;x_j,z)\,+\,\right. 
\gamma_2(x_i,y_j; z,y_i) 
-\left.\gamma_2(x_i,y_j; x_j,y_i)\right] \nonumber
\nonumber
\end{eqnarray}
In \eq{NG1} we have omitted terms proportional to $\gamma_3$.
Substituting $$\gamma_1\lb{x}, {y};Y\rb\,\equiv\,N\lb {x}, {y};Y\rb$$ and 
$$\gamma_2\lb {x}_1, {y}_1; {x}_2, {y}_2;Y\rb\,=\,N\lb {x}_1, {y}_1;Y\rb\,
\,N\lb {x}_2, {y}_2;Y\rb\,+\,\,C\lb  {x}_1, {y}_1; {x}_2, {y}_2;Y \rb $$ 
we obtain assuming  the correlation function $C$ is small,  $C < N^2$:
 \begin{eqnarray} \label{NG3}
\frac{\partial\,N\lb {x}, {y};Y\rb}{\partial\,Y}&=&\bas\int\,d^2z\,K\lb 
 {x}, {y}; 
 {z} \rb\,\lb   {\cal N}\lb  {x},  {y}; {z}\rb
-\,N\lb  {x},  {z} \rb\,N \lb  {x},  {z} 
\rb -\,C \lb  {x},  {z};  {y} ,  {z};Y \rb  \rb =  \nonumber \\
&=&\,\bas\,\int\,d^2\,z\,\,K\lb
 {x}, {y}; {z} \rb\,\,\lb \tilde{\cal{N}}\lb  {x},  {y}; {z}\rb\,\,-\,\,C\lb 
 {x},  {z};  {y} ,  {z};Y \rb  \rb; 
\end{eqnarray}
where we define
\beq \label{NCAL}
{\cal N}\lb  {x},  {y}; {z}\rb\,\equiv\,N\lb  {x},  {z} \rb \,+\,N\lb  {y}, 
 {z} \rb\,-\,N\lb  {x},  {y} \rb
\eeq
$$ 
\tilde{\cal {N}}\lb  {x},  {y}; {z}\rb\,\equiv\,{\cal N}\lb  {x}, 
 {y}; {z}\rb\,-\,
\,N\lb  {x},  {z} \rb\,N \lb  {x},  {z}\rb
$$
$$
 {\,\cal C}\lb  {x}_1,  {y}_1, {z}; {x}_2,  {y}_2;Y\rb\,\,\equiv\,\,
C\lb   {x}_1, {z};  {x}_2,  {y}_2;Y \rb \,+\,C\lb   {y}_1, {z};  {x}_2, 
 {y}_2;Y \rb, \,\,-\,\,C\lb   {x}_1, {y}_1;  {x}_2,  {y}_2;Y \rb
$$
The equation for $\gamma_2$ becomes an equation for $C$
\beq  \label{NG4}
\frac{\partial\,C \lb  {x}_1  {y}_1;  {x}_2 ,  {y}_2;Y \rb}{\partial\,Y} \,\,=
\eeq
$$
\,\,\frac{\bas}{2}\, \int\,d^2\,z\, \lb K\lb  {x}_1 {y}_1; {z} \rb 
  {\cal{C}} \lb  {x}_1,  {y}_1,  {z};  {x}_2 ,  {y}_2;Y  \rb  \,+\, K\lb 
 {x}_2 {y}_2; {z} \rb\,\,{\cal{C}} 
\lb 
 {x}_1 ,  {y}_1 ;  {x}_2  {y}_2,  {z}; Y \rb \rb
$$
$$
+\,\,\frac{\bas}{2\,N^2_c}\,\lb \int\,d^2\,z\,\lb  K\lb  {y}_1, {x}_2; {z} \rb
\,{\cal{C}} \lb  {y}_1, {x}_2,  {z};  {x}_1, {y}_2; Y \rb\,\,+\,\,
  \,K\lb  {x}_1, {y}_2; {z} \rb
{\cal{C}} \lb  {x}_1, {y}_2,  {z};  {x}_2, {y}_1; Y \rb \rb \right.  
$$
$$
\left.
+ \,\,\int\,d^2\,z\,  K\lb  {y}_1, {x}_2; {z} \rb\, \lb
\,N\lb  {x}_1, {y}_1 \rb \,+\,N\lb  {x}_2, {y}_2 \rb \,-\, N\lb 
 {x}_1, {y}_2 \rb\,\rb \,\lb \tilde{\cal {N}}\lb  {y}_1, {x}_2; {z} \rb
\,+\, \tilde{\cal{N}}\lb  {x}_1, {y}_2; {z} \rb \rb \rb
$$

It is interesting to notice that \eq{NG4} can be reduced to the equation of Bartels, Lipatov and 
Vacca \cite{BLV}. Indeed, we can introduce a new function $\Delta C$:
\beq \label{DELC}
\Delta C \lb  {x}_1,  {y}_1; {x}_2,  {y}_2;Y\rb\,\,\,=
\eeq
$$
 C \lb  {x}_1, 
 {y}_1; {x}_2,  {y}_2;Y\rb\,\,-\,\,\frac{1}{N^2_c}\,\lb N\lb {x}_1, {y}_1;Y \rb 
\,+\,
N\lb {x}_2, {y}_2;Y \rb\,-\,N\lb {x}_1, {y}_2;Y \rb \rb^2
$$
Using \eq{NG3} we can reduce the set of \eq{NG3} and \eq{NG4} to a different set of equations, 
namely,
\beq \label{NG5}
\frac{\partial\,N\lb {x}, {y};Y\rb}{\partial\,Y}\,\,=
\eeq
$$
\bas\,\int\,d^2\,z\,\,K\lb
 {x}, {y}; {z} \rb\,\,\lb {\cal{N}}\lb  {x}, {y}; {z}\rb\,\,-\,\,\,N\lb  {x}, 
 {z} \rb\,N \lb  {x},  {z}\rb 
-\frac{1}{2\,N^2_c}{\cal{N}}^2\lb  {x}, {y}; {z}\rb \,\,+\,\,\Delta C\lb
 {x},  {z};  {y} ,  {z};Y \rb  \rb ;
$$
\beq \label{NG6}
\frac{\partial\,\Delta C\lb
 {x},  {z};  {y} ,  {z};Y \rb}{\partial\,Y}  \,\,=\,\,
\eeq
$$
\frac{\bas}{2}\, \int\,d^2\,z\, \lb \, K\lb
 {x}_2, {y}_2; {z} \rb\,\,\Delta {\cal{C}}\lb  {y}_1, {x}_2,  {z}; 
 {x}_1, {y}_2; Y \rb\,\,+\,\,
  \,K\lb  {x}_1, {y}_2; {z} \rb
\,\,\Delta {\cal{C}} \lb  {x}_1, {y}_2,  {z};  {x}_2, {y}_1; Y \rb \rb\,+
$$
$$
\,+\,\,\frac{\bas}{2\,N^2_c}\,\int\,d^2\,z\,\lb  K\lb  {y}_1, {x}_2; {z} \rb
 N\lb  {y}_1,  {z};  Y \rb\,\,N\lb  {x}_2,  {z};  Y \rb \,\,+\,\,
  \,K\lb  {x}_1, {y}_2; {z} \rb
N\lb  {y}_1,  {z};  Y \rb\,\,N\lb  {x}_2,  {z};  Y \rb \rb
$$
These two equations are the same as were proposed by Bartels, Lipatov and Vacca \cite{BLV}.
Our derivation suggests also a physical meaning of the modified Balitsky-Kovchegov equation (see 
\eq{NG5}). The  Balitsky-Kovchegov equation is a mean field approximation while \eq{NG5} takes 
into account the correlation related to possibility for  grouping of two dipoles in a different 
way with suppressed probability.  Therefore, it plays a role of Fock term in Hartree-Fock 
approach, which is a natural next step in the mean field approach. $\Delta C$ is a real dynamic 
correlations which as one can see from \eq{NG6} grows with energy. We have
neglected terms of the order  $\Delta C N$ in comparison with $N^2$ -term. Therefore, we can 
trust the equations only for $\Delta C \leq N$. For higher energies we need to develop a more 
general approach.

\section{Conclusions and Discussion}

In this paper we have extended our linear operator approach applied to dipole evolution.
The evolution kernel $\chi$ can be viewed as a ``Hamiltonian'' of the evolution. It is constructed
in terms of dipole creation ad annihilation operators. By introducing the recombination vertex
$\Gamma \lb 2 \rightarrow 1 \rb$, the evolution operator has been promoted to a fully quantum 
two dimensional  field theory of interacting dipoles (Pomerons).

The main results of this paper are  \eq{Z}, \eq{rho}, \eq{N5}, \eq{chi23} and \eq{gamma23}
supplemented by the explicit expressions for the vertices  
$\Gamma \lb 
2 \rightarrow 1 \rb$ and $ \Gamma \lb 2 \rightarrow 3 \rb$ which both are 
proportional to the 
second functional derivative with respect to $u_i$. Our  
approach is an extension beyond the 
Balitsky one \cite{B}, based on the Wilson loops, as well as beyond the 
Color Glass Condensate  approach (JIMWLK equation \cite{MV,JIMWLK}). 
Though  the 
JIMWLK equation takes into account all  $1/N^2_c$ corrections, and which are
only partially accounted for by the vertex $\Gamma(2\rightarrow 3)$, they do
not include the recombination vertex $\Gamma(2\rightarrow1)$ which is a major step
beyond this equation.

\begin{figure}[htbp]
\begin{center}
\epsfig{file= 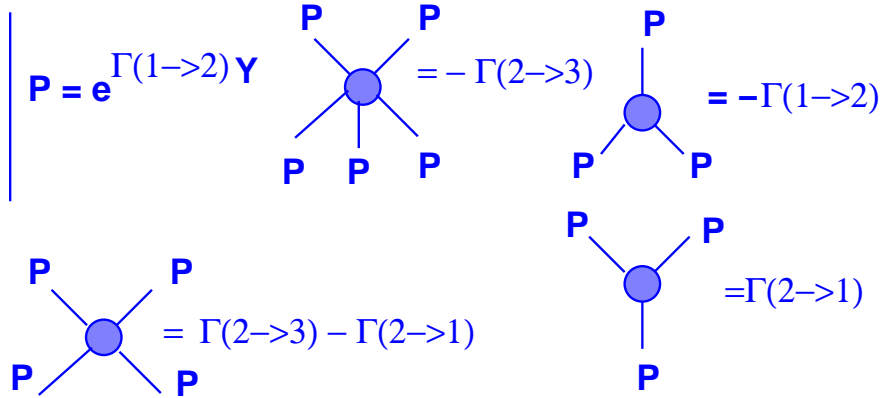,width=140mm}
\end{center}
\caption{\it Pomeron interactions described by \eq{C2}.}
\label{pom}
\end{figure}

We have accounted for  dynamical correlations that 
stem  from possibility of merging of two BFKL Pomerons. 
It is  illustrative to  consider a 
simple 
toy model in which we assume  that interactions 
do not depend on the dipole sizes (see 
Refs. \cite{MUDM,LL,KOLE} for details).
The master functional equation (see \eq{Z}) for this model 
degenerates into ordinary equation 
in partial derivatives
\beq \label{C1}
\frac{\partial\,Z}{\partial Y}\,
=\,- \Gamma(1 \rightarrow 2)\,u(1 - 
u) 
\,\frac{\partial\,Z}{\partial u}\,+\,\Gamma(2 \rightarrow 1)\,\,u(1 - 
u)\,\,\frac{\partial^2\,Z}{(\partial u)^2}\,+\, \Gamma(2 \rightarrow 3)
u\,(1 - 
u)^2\frac{\partial^2\,Z}{(\partial u)^2}
\eeq
We can introduce a generating function for the scattering 
amplitude using the relation \cite{LL}
\beq \label{ZN}
N(Y;u)\,\,=\,\,1\,\,-\,\,Z(Y; 1 - \gamma)
\eeq
To obtain the scattering amplitude we need to replace $\gamma$ in \eq{ZN},  by the amplitude 
of interaction of a dipole with the target. 
For $N$ \eq{C1} can be rewritten in the form:
\beq \label{C2}
\frac{\partial\,N}{\partial Y}\,
=\, \Gamma(1 \rightarrow 2)\,\,\gamma(1 - 
\gamma)\frac{\partial\,N}{\partial \gamma}
\,+\,\Gamma(2 \rightarrow 1)\,\,\gamma (1 - \gamma)\,\frac{\partial^2\,N}{(\partial 
\gamma)^2}\,\,+\,\, \Gamma(2 \rightarrow 3)
\gamma^2\,(1 -\gamma)\frac{\partial^2\,N}{(\partial \gamma)^2}
\eeq
if $\gamma$ is small we can reduce \eq{C2} to a simpler equation
\beq \label{C3}
\frac{\partial\,N}{\partial Y}\,\,=\,\,\Gamma(1 \rightarrow 2)\,\,\gamma 
\,\frac{\partial\,N}{\partial \gamma}\,.
\eeq
The solution of \eq{C3} is a Pomeron with the intercept $\Gamma(1 \rightarrow 2)$:
\beq \label{C4}
N\,\,=\,\,\gamma e^{ \Gamma(1 \rightarrow 2) Y}
\eeq
The rest of the terms in \eq{C2} are responsible for Pomeron interactions
(see \fig{pom}).
 
As we see from \fig{pom} two vertices  $\Gamma(2 \rightarrow 1)$ and  $\Gamma(2 \rightarrow 3)$
are responsible for different processes of Pomeron interaction. At first sight $\Gamma(2 \rightarrow 
1)$ is much smaller than  $\Gamma(2 \rightarrow 3)$ and can be neglected. However, we can make such a 
conclusions only if we will find out what value of $u $ ( or $\gamma$) are essential for high 
energies. Therefore,  the vertex $\Gamma(2 \rightarrow 1)$ can be still relevant in certain
kinematic domains. To answer this question we need a detailed analysis of \eq{N5}  and \eq{gamma23}
which is beyond the scope of this paper (this question is addressed in  Ref. \cite{Levin2005}).

\begin{figure}[htbp]
\begin{center}
\epsfig{file= 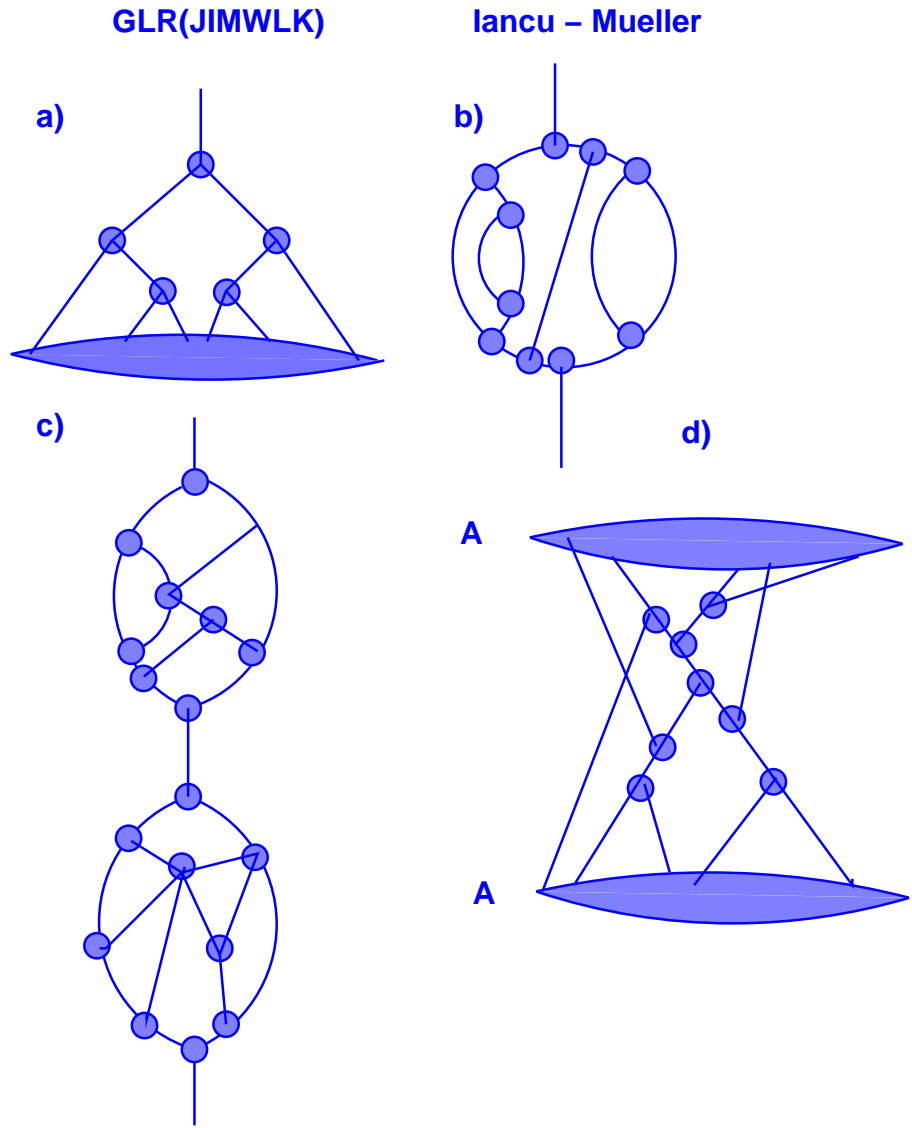,width=120mm}
\end{center}
\caption{\it The typical Pomeron diagrams for interactions described by \eq{C2}. \fig{pomint}-a 
describes the GLR approach \cite{GLR} which for \eq{C2} coincide with the Balitsky-JIMWLK 
approach \cite{JIMWLK,B}; 
\fig{pomint}-b corresponds to Iancu-Mueller approach\cite{IM} which suggests the way out of the 
JIMWLK approach and can be justified in limited region of energy. \fig{pomint}-c shows the general 
type of the diagrams that can be summed in the framework of the approach based on \eq{C2}.  
\fig{pomint}-d are diagrams that we need to sum for nucleus-nucleus interaction at high energy. }
 \label{pomint} 
\end{figure}

\fig{pomint} presents some examples of Pomeron diagrams which correspond to different 
approaches that has been discussed in the past:  the GLR equation \cite{GLR} (see \fig{pomint}-a)
 which, in our approximation, coincides with  the BK  \cite{B,K}    and JIMWLK 
\cite{JIMWLK} equations;    the Iancu-Mueller approach \cite{IM} (see \fig{pomint}-b).   \fig{pomint}-c 
shows a typical diagram that can be incorporated using \eq{C2}. Finally, in  \fig{pomint}-d 
we  plot the diagrams that one needs to sum in order to reliably consider nucleus - nucleus 
interactions. In 
general such  diagrams are difficult to sum, but we have an experience that in the simple 
model of \eq{C2}, this summation can be performed \cite{Braun,BOGLM}.

It is important to stress that by introducing the vertex $\Gamma(2\,\rightarrow \,3)$, we have 
taken into account only  the leading $N_c$ corrections. For $n$ dipole densities with 
$n>2$
we should have color correlations which cannot be presented in the dipole basis. We believe,
however, that these correlations are of no significance at high energies.

The importance of correlations have been already noticed in 
Refs.\cite{BALE,LALE,IM}.
\eq{C2} illustrates a complexity of the problem since even this oversimplified  
equation has not been solved.  
The expansion in correlations allows to shrink the infinite hierarchy of equations
to a system of two coupled equations.  This reduction provides  a method for
estimating  importance of both the $N_c$ correlations and Pomeron loops.
We demonstrated  that correlations should be essential 
at high energies and suggested a consistent  approach to take them into account.

In this paper we have considered a merging process of  two Pomerons into one only. 
In general, there exist higher order processes accounting for a possibility of many Pomerons
merging into one. A formal resummation of these processes has been reported in recent Ref. \cite{KL1}
and also in Ref. \cite{SMITH}.

We hope that we propose the simplest way of dealing with the Pomeron loops which is equivalent to the 
reggeon calculus for BFKL Pomeron but has an advantage of clear probabilistic interpretation in the 
rest frame of one of the colliding particles. We hope that  clarification of all assumptions in our 
approach will 
lead us to deeper and more transparent  understanding of physics in the saturation domain.

Finally, let us comment on two recent papers
\cite{MUSHWO,IT1}  which appeared practically simultaneously with ours and
contain features  close to those presented here.
In fact the formal expression for the vertex $\Gamma (2 \rightarrow 1)$ (\eq{ga21})
is identical to the  ones of Refs. \cite{MUSHWO,IT1}. 
This equivalence has been proven in a later Ref. \cite{KL3}.
 The main difference is that we have extended the method. Apart from giving the
 formal  expression for the vertex $\Gamma (2 \rightarrow 1)$, 
 we also introduced a formalism needed for its
 evaluation (see Appendix). At the end, we were able to obtain a first analytical evaluation of the vertex
 bringing it to the level ready for computer simulations (\eq{A13}).  
The diagonal transition $2\rightarrow2$, which  guarantees probability conservation, vanishes if the
exact expression for the vertex $\Gamma (2 \rightarrow 1)$ is used. Generally this term should not be 
neglected if an approximate vertex is used for  practical applications. 
In addition we have included the vertex $\Gamma (2 \rightarrow 3)$ in our consideration.

\section*{Acknowledgments:}
We want to thank Asher Gotsman, Alex Kovner and Uri Maor for very useful 
discussions on the subject 
of this paper. 
We are most grateful to  Nestor Armesto for pointing to several misprints 
in our manuscript and for very valuable comments.  
We are very grateful to our referee  whose comments and constructive criticism 
helped us to considerably improve this paper.
This research was supported in part  by the Israel Science Foundation,
founded by the Israeli Academy of Science and Humanities.

\appendix
\begin{boldmath}
\section*{Appendix A:\,\, Calculation of $\Gamma \lb 2 \rightarrow 1 \rb$.}
\end{boldmath}
 \renewcommand{\theequation}{A.\arabic{equation}}
\setcounter{equation}{0}
In this appendix we find the solution to \eq{G213}. Our approach is based on the main 
properties of the BFKL kernel which have been studies in details in Refs. \cite{LI,NP}. 
First, we rewrite \eq{G21G} in the form of the contour integral over $h$  \cite{LI,NP}, 
namely,  
\beq \label{A0}
\gamma^{BA} \lb x,y; x',y' \rb\,\,=\,\,\frac{1}{2}\,\{\tilde\gamma_{BA} \lb x,y; x',y' \rb 
\,\,-\,\,\tilde\gamma_{BA} \lb x,y; y',x' \rb \}
\eeq
where
\beq \label{A1}
\tilde\gamma_{BA} \lb x,y; x',y' \rb \,\,=
\eeq
$$
=\,\,\int^{a + i \infty}_{a - i\infty}\,\,\frac{d\,h}{2
\,\pi\,i\,h^3}\,\tilde\gamma_{BA}\lb h; x,y; x',y' \rb\,\,=\,\,
\frac{\bas^2}{16\,N^2_c}\,\int^{a + i \infty}_{a - i 
\infty}\,\,\frac{d\,h}{2 \,\pi\,i}\,\frac{1}{h^3}\,\lb \frac{( {x} - 
 {y})^2\,( {x}^{\,'} -  {y}^{\,'})^2}{( {x} -  {x}^{\,'})^2\,( {y} - 
 {y}^{\,'})^2}\rb^h
$$
where $x,y,x'$ and $y'$ is the coordinate of quarks and antiquarks in the interacting dipoles 
with the size $r = x - y$ and $ r' = x' - y'$. Introducing complex numbers instead of vectors  
$x = x_1 + i x_2$ and $x^*=x_1 - i x_2$  $( x \,x^*\,=\,( {x})^2 )$,  we can rewrite 
$\tilde\gamma_{BA}\lb h; x,y; x',y' \rb $ in the form:
\beq \label{A2}
\tilde\gamma_{BA}\lb h; x,y; x',y' \rb\,\,=\,\frac{\bas^2}{16\,N^2_c}\,\lb
\frac{(x - y)\,(x' - y')}{(x - x')\,(y - y')}\rb^h\,\,\lb \frac{(x - y)^*\,(x' 
- y')^*}{(x - x')^*\,(y - 
y')^*}\rb^h\,\equiv\,\frac{\bas^2}{16\,N^2_c}\,\,\gamma_{BA}^h\,\times\,\gamma^{h\,*}_{BA}
\eeq
At first sight  \eq{A0} does not lead to  \eq{G21G}. Indeed it gives 
\beq \label{A01}
\gamma^{BA} \lb x,y; x',y' \rb\,\,=
\eeq
$$
=\,\,\frac{\bas^2}{32\,N^2_c} \,\ln \lb 
\frac{( {x} -
 {y}^{\,'})^2\,( {y} -  {x}^{\,'})^2}{( {x} -  {x}^{\,'})^2\,( {y} 
-
 {y}^{\,'})^2}\rb\, \,\,\ln \lb
\frac{( {x} -
 {y})^4\,( {x}^{\,'} -
 {y}^{\,'})^4}{( {y} -  {x}^{\,'})^2\,( {x} -  {y}^{\,'})^2\,( {y}
-
 {y}^{\,'})^2\,\,( {x} -  {x}^{\,'})^2}\rb
$$
The replacement of the Born amplitude \eq{G21G} by \eq{A0} is a major step for what follows
and has to be justified.  We refer here to the work of Lipatov \cite{LI} who showed that
 the Born amplitude could be written in the form of 
\eq{A0} (see Eq.110 of the first paper in Ref.  \cite{LI}). The main idea of Ref. \cite{LI} is that  
two expressions  \eq{A0} and  \eq{G21G} lead to the very same results if  used for calculations of  physical 
observables 
(for example $\gamma-\gamma$ scattering). Both expressions  satisfy \eq{DDBA} and hence they 
differ by a function $\xi$, which does not depend on one of the coordinates $x$ (or $y$). Lipatov showed that,
thanks to the properties of the impact factor $\Phi(x,y,q)$ 
(see Eq. 109 in Ref.  \cite{LI}),
 the integral over $x$ (or $y$) of 
the impact factor convoluted with $\xi$ vanishes.
 This property of the impact factor implies that a function, which does not depend on one of the 
coordinates, gives zero contribution to any  physical process.  Moreover, the well known  BFKL Green function 
\cite{BFKL}  was calculated using \eq{A0} as initial condition. To be consistent with the use
of the BFKL kernel,  \eq{A0} has to be taken as the Born  approximation. We will see below that
this replacement allows us to evaluate the vertex $\Gamma(2\rightarrow1)$ (\eq{ga21}).

In what  follows we deal with the first term in \eq{A0} but it is a trivial algebraic exercise 
to obtain a result for the full Born amplitude of \eq{A0}.

The r.h.s. of \eq{G213} we rewrite, using \eq{A1} and \eq{A2} in the form
$$
\bas\,\int^{a + i \infty}_{a - i\infty}\,\,\frac{d\,h_1}{4
\,\pi\,i}\,\frac{1}{h^3_1}\,\,\,\int^{a + i \infty}_{a - i\infty}\,\,\frac{d\,h_2}{4
\,\pi\,i}\,\frac{1}{h^3_2}\,\int \,\,d z\,\,d z^*\,\lb \frac{x -y}{(x - z)\,(y - 
z)}\rb\,\cdot\,\lb 
\frac{x -y}{(x - z)\,(y - z)}\rb^*
$$
\beq \label{A3}
\,\,\tilde\gamma_{BA}\lb h_1; x,z; x_1,y_1 
\rb\,\,\,\tilde\gamma_{BA}\lb h_2; z,y; x_2,y_2 \rb
\eeq

The integrals over $z$ and $z^*$ can be computed using formula {\bf 3.211}, {\bf 9.182}(1) and 
{\bf 9.183}(1) of Ref. \cite{RY}. Indeed, \eq{A3} can be rewritten as follows
\beq \label{A4}
\frac{\bas^3}{(16\,N^2_c)^2}\,\int^{a + i \infty}_{a - i\infty}\,\,\frac{d\,h_1}{4
\,\pi\,i}\,\frac{1}{h^3_1}\,\,\,\int^{a + i \infty}_{a - i\infty}\,\,\frac{d\,h_2}{4
\,\pi\,i}\,\frac{1}{h^3_2}\,\lb I\,\,I^* \rb
\eeq

Using the notation $x - y =r$ and $x - z =r'$, we have
\beq\label{A50}
 I\,\,=\,\,\int\,d r' \frac{r}{(r - r')\,r'} \,
\lb \frac{r'\,(x_1 - y_1)}{(x - x_1)\,(x-r' - y_1)}\rb^{h_1}
\lb \frac{(r' - r)\,(x_2 - y_2)}{(y - y_2)\,(x -r' - x_2 )}\rb^{h_2}
\eeq

We calculate $I$ dividing the integration over $r'$ in three regions\footnote{We take the 
integral
over $r'$ along the real axis. The final answer we obtain by analytic continuation of all
integrals into complex plane for all variable.},
 namely,
\beq \label{I}
I\,\,=\,\,I_1 \lb -\infty\, <\, r'\, <\,0 \rb\,+\,I_2 \lb 0\, < r'\, < \,r \rb\,+\,I_3\lb r\, < 
r'\, < \,\infty \rb
\eeq
 Let us  first compute the integral from the second region 
\begin{eqnarray}
I_2\,\,&=&\,\,\int_0^r\,d r' \frac{r}{(r - r')\,r'} \,
\lb \frac{r'\,(x_1 - y_1)}{(x - x_1)\,(x -  r' - y_1 )}\rb^{h_1}
\lb \frac{(r' - r)\,(x_2 - y_2)}{(y - y_2)\,(x - r' -x_2 )}\rb^{h_2}
\,\,= \nonumber\\
&=&\,\,\lb \frac{(x_1 - y_1)\,(x -y)}{(x - x_1)\,(x - y_1)} \rb^{h_1}\,\lb \frac{(x_2 - 
y_2)\,(x -y)}{(y - y_2)\,(x - x_2)} \rb^{h_2}\,\, \label{A52}\\
 &\times &\,\,\frac{\Gamma(h_1)\,\Gamma(h_2)}{\Gamma(h_1+h_2)}\,\,F_1 \lb 
h_1,h_1,h_2,h_1+h_2,\frac{(x-y)}{(x - y_1)},\frac{(x-y)}{(x - x_2)} \rb \nonumber \\
&=&\lb \frac{(x_1 - y_1)(x - y)}{(x - x_1)\,(y - y_1)} \rb^{h_1}\,\lb \frac{(x_2 -
y_2)(x -y)}{(x
-
x_2)\,(y  - y_2)} \rb^{h_2}\,\, \label{A53} \\
&\times& \frac{\Gamma(h_1)\,\Gamma(h_2)}{\Gamma(h_1+h_2)}
\,{}_2F_1\lb h_1,h_2,h_1 + h_2,\frac{(x - y)\,(x_2 - y_1)}{(x - x_2)(y - y_1)} \rb
\nonumber \\
 & \rightarrow & \frac{(h_1+h_2)}{h_1\,h_2}\, 
\lb \frac{(x_1 - y_1)(x - y)}{(x - x_1)\,(y - y_1)} \rb^{h_1}\,\lb \frac{(x_2 -
y_2)(x -y)}{(x
-
x_2)\,(y  - y_2)} \rb^{h_2}\label{A54}
\end{eqnarray}
 \eq{A52} was obtained  using  {\bf 3.211} of Ref. \cite{RY}.
Obtaining \eq{A54} we take into account that only small values of $h_1$ and $h_2$   
 will contribute to the integral of \eq{A3}.

In \eq{A52} and \eq{A54} $F_1$ and ${}_2F_1 \equiv F$ denote the hypergeometric functions (see
formula {\bf 9.10} and {\bf 9.180}(1) in Ref. \cite{RY}).

$I_3$ is the integral of \eq{A50} for $ r' >r$, namely,

\begin{eqnarray}
 I_3\,\,&=&\,\,\int_{r}\,d r' \frac{r}{(r - r')\,r'} \,
\lb \frac{r'\,(x_1 - y_1)}{(x - x_1)\,(x - r'  - y_1)}\rb^{h_1}
\lb \frac{(r - r')\,(x_2 - y_2)}{(y - y_2)\,(x -r' - x_2)}\rb^{h_2}\,=\nonumber\\
&=&\,\,\lb \frac{(x_1 - y_1)}{(x - x_1)} \rb^{h_1}\,\lb \frac{(x_2 -
y_2)}{(y - y_2)} \rb^{h_2}\,\, \label{A61}\\
 &\times &\,\,\frac{\Gamma(1)\,\Gamma(h_2)}{\Gamma(1 +h_2)}\,\,F_1 \lb
1,h_1,h_2,1 +h_2,\frac{(y - y_1)}{(x -y)},\frac{(y - y_2)}{(x -y)} \rb \nonumber \\
&\rightarrow&\,\, \frac{1}{h_2} \lb \frac{(x_1 - y_1)}{(x - x_1)} \rb^{h_1}\,\lb \frac{(x_2
-
y_2)}{(y - y_2)} \rb^{h_2} \label{A63}
\end{eqnarray}

In \eq{A63} we found the limit at small values of $h_2$ which contribute to the integral of 
\eq{A3}. The third integral is equal to
\begin{eqnarray}
 I_1\,\,&=& \,\,\int_{-\infty}^0\,d r' \frac{r}{(r - r')\,r'} \,
\lb \frac{r'\,(x_1 - y_1)}{(x - x_1)\,(x-r' -y_1 )}\rb^{h_1}
\lb \frac{(r - r')\,(x_2 - y_2)}{(y - y_2)\,(x -r' -x_2 )}\rb^{h_2}\,=\nonumber\\
&\rightarrow&\,\, \frac{1}{h_1} \lb \frac{(x_1 - y_1)}{(x - x_1)} \rb^{h_1}\,\lb \frac{(x_2
-
y_2)}{(y - y_2)} \rb^{h_2} \label{A063}
\end{eqnarray}

Substituting \eq{A54} and \eq{A63} into \eq{A4} we finally obtain the result for the r.h.s. 
of \eq{G213}, namely,
\beq \label{A7}
\frac{\bas^5}{(16\,N^2_c)^2}\,\int^{a + i \infty}_{a - i\infty}\,\,\frac{d\,h_1}{4
\,\pi\,i}\,\frac{1}{h^3_1}\,\,\,\int^{a + i \infty}_{a - i\infty}\,\,\frac{d\,h_2}{4
\,\pi\,i}\,\frac{1}{h^3_2}\,\lb \frac{(h_1+h_2)}{h_1\,h_2}\rb^2\,\times
\eeq
$$
 \lb\lb
 \frac{(x_1 - y_1)(x - y)}{(x - x_1)\,(y - y_1)} \rb^{h_1}\,\lb \frac{(x_2 -
y_2)(x -y)}{(x
-
x_2)\,(y  - y_2)} \rb^{h_2} \,+\,
 \lb \frac{({x}_1 - {y}_1)}{({x} - {x}_1)} \rb^{h_1}\,\lb 
\frac{({x}_2 - {y}_2)}{({y} - {y}_2)} \rb^{h_2} \rb \,\times
$$
$$
\lb \lb 
\frac{(x_1 - y_1)(x - y)}{(x - x_1)\,(y - y_1)} \rb^{h_1}\,\lb \frac{(x_2 -
y_2)(x -y)}{(x
-
x_2)\,(y  - y_2)} \rb^{h_2} \,+\,
 \lb \frac{({x}_1 - {y}_1)}{({x} - {x}_1)} \rb^{h_1}\,\lb
\frac{({x}_2 - {y}_2)}{({y} - {y}_2)} \rb^{h_2} \,\rb^*
$$
The integrals over $h_1$ and $h_2$ can be evaluated 
but we postpone this until we work out the action of Laplacians (\eq{ga21}).

As was noticed in Section 3, the Born amplitude in the form of \eq{A1} as well as of 
\eq{G21G} satisfy the following equation
\beq \label{A8}
\Delta_x\,\Delta_y\,\tilde\gamma_{BA} \lb x,y; x',y' \rb\,\,\equiv\,\frac{d}{d 
x}\,\frac{d}{d 
x^*}\,\frac{d}{d y}\,\frac{d}{dy^*}\,\tilde\gamma_{BA} \lb x,y; x',y' \rb\,\,\,
\eeq
$$
=\,\,\as^2 \lb
\delta^{(2)}\lb  {x} -  {x}^{\,'}\rb \delta^{(2)}\lb  {y} -  {y}^{\,'}\rb
\,+\,\delta^{(2)}\lb  {x} -  {y}^{\,'}\rb \delta^{(2)}\lb  {y} - 
 {x}^{\,'}\rb \rb
$$
Thus $\Gamma(2\rightarrow1)$ (see \eq{ga21}) is obtained by
applying operator $\Delta_x\,\Delta_y$ to \eq{A7} and multiplying by 
$N_c^2/\bar\alpha_s^2$.
The observation which helps to simplify the calculation is the following:
\beq \label{DEL1}
\frac{d}{d x}\,\frac{d}{d y}\,I_2 \,\,=\,\, \lb \frac{1}{h_1} + \frac{1}{h_2}\rb 
\,\lb \lb h_1 \,+ \,h_2 \rb\,\frac{1}{(x - y)^2}\,\,\,+\,\,O(h^2) \rb\,I_2\,,
\eeq
where we can neglect terms that are proportional to $h_1^2$ or $h^2_2$, as the 
dominant contribution  to \eq{A8} stems from the region of small $h$'s.
Similarly the contribution originating from the integrals $I_1$ and $I_3$ can be 
neglected since
\beq \label{DEL2}
\frac{d}{d x}\,\frac{d}{d y}\,I_1 \,\,\propto\,\,h^2_1\frac{1}{h_1}\,
\eeq
Taking into account \eq{DEL1} and \eq{DEL2} we obtain
\beq \label{A10}
\Gamma_{2\rightarrow 1} \lb (x_1,y_1) + (x_2,y_2)\,\rightarrow\,(x,y) \rb \,\,=\,\,
\frac{N_c}{\pi\,\alpha_s}\,\int^{a + i \infty}_{a - i\infty}\,\,\frac{d\,h_1}{4
\,\pi\,i}\,\frac{1}{h^3_1}\,\,\,\int^{a + i \infty}_{a - i\infty}\,\,\frac{d\,h_2}{4
\,\pi\,i}\,\frac{1}{h^3_2}\,\,\frac{(h_1+h_2)^4}{h^2_1\,h^2_2}
\eeq
$$
\frac{1}{( {x} -  {y} )^4 }\,\,\tilde{\gamma}_{BA}\lb h_1; x,y;
x_1,y_1 \rb\,\,
\tilde{\gamma}_{BA}\lb h_2; x,y; x_2,y_2\rb\
$$
Now we can easily evaluate
the remaining
integrals over $h_1$ and $h_2$. The result can be written in the most 
economic form introducing a new variable:
\beq \label{A11}
R\lb x,y; x',y' \rb \,\,=\,\, \frac{( {x} -  {y})^2\,\,( {x}^{\,\,'} - 
 {y}^{\,\,'})^2}{( {x} - {x}^{\,\,'})^2\,\,( {y} - {y}^{\,\,'})^2}
 \eeq
The vertex reads
\beq \label{A12}
\Gamma_{2\rightarrow 1} \lb (x_1,y_1) + (x_2,y_2)\,\rightarrow\,(x,y) \rb \,\,=\,\,
\frac{\bas^3}{(32 \,\pi)^2\,\,N^2_c}\,\,\frac{1}{( {x} -  {y} )^4 } \,\,
\lb \frac{1}{24}\,\ln^4 R\lb x,y; x_1,y_1 \rb\,+ \right.
\eeq
$$
\left.
\,\frac{2}{3}\,\ln^3 R\lb x,y; 
x_1,y_1 
\rb\ln R\lb x,y; x_2,y_2 \rb\,+\,\frac{3}{2}\,\ln^2 R\lb x,y; x_1,y_1
\rb\ln^2 R\lb x,y; x_2,y_2 \rb\,+ \right.
$$
$$
\left.
\,\,+\,\, \frac{2}{3}\, \ln R\lb x,y; x_1,y_1
\rb\ln^3 R\lb x,y; x_2,y_2 \rb\,+\,\frac{1}{24}\,\ln^4 R\lb x,y; x_2,y_2 \rb \rb
$$
So far, we evaluated the contribution of the first term ($\gamma_{BA}^{(1)}$) of
the full Born amplitude of \eq{A0}. Having added the second term we end up with the
final expression for the vertex:
\begin{eqnarray} 
\Gamma_{2\rightarrow 1} \lb (x_1,y_1) + (x_2,y_2)\,\rightarrow\,(x,y) \rb \,=\,
\frac{\bas^3}{2\,(32 \,\pi)^2\,N^2_c}\,\,\frac{1}{( {x} -  {y} )^4 } \,
\ln\frac{R\lb x,y;x_1,y_1\rb}
{R\lb x,y;y_1,x_1\rb}\,\ln\frac{R\lb x,y;x_2,y_2\rb}{R\lb 
x,y;y_2,x_2\rb}\times
\nonumber 
\end{eqnarray}
\begin{eqnarray}
&\lb 
\,\frac{2}{3}\,[\ln^2 R\lb x,y;x_1,y_1\rb\,\,+\,\,\ln R\lb x,y;x_1,y_1\rb\,\ln R\lb 
x,y;y_1,x_1\rb\,\,+\,\,\ln^2 R\lb x,y;y_1,x_1\rb]\,\,+ \right.
\nonumber \\
&\nonumber \\
&
\left.
\,+\,\frac{3}{2}\,\ln( R\lb x,y; x_1,y_1\rb\,R\lb x,y; y_1,x_1\rb)
\,\ln( R\lb x,y; x_2,y_2\rb R\lb x,y; y_2,x_2\rb)\, + \right.
\label{A13} \\
&\nonumber \\
& \left.
\,\,+\,\,\frac{2}{3}\,[\ln^2 R\lb x,y;x_2,y_2\rb\,\,+\,\,\ln R\lb x,y;x_2,y_2\rb\,\ln R\lb
x,y;y_2,x_2\rb\,\,+\,\,\ln^2 R\lb x,y;y_2,x_2\rb] 
 \rb \nonumber
\end{eqnarray}

\end{document}